\documentclass{emulateapj} 
\usepackage{apjfonts}
\usepackage{epsfig}
\usepackage{xspace}
\usepackage{natbib}
\usepackage{hhline}
\usepackage{amsmath}
\usepackage{multirow}

\newcommand{\Chandra}{{\it Chandra}\xspace}
\newcommand{\champlane}{ChaMPlane\xspace}
\newcommand{\champ}{ChaMP\xspace}
\newcommand{\wavdetect}{{\it wavdetect}\xspace}
\newcommand{\sherpa}{{\it sherpa}\xspace}

\newcommand{\sS}[1]{\mbox{$\rm{}^{#1}$}}
\newcommand{\Ss}[1]{\mbox{$\rm{}_{#1}$}}

\newcommand{\nH}{\mbox{$N_{\mbox{\scriptsize H}}$}\xspace}
\newcommand{\nHx}{\mbox{$N_{\mbox{\scriptsize H}}$}}
\newcommand{\Sx}{\mbox{$S_X$}\xspace}
\newcommand{\Hx}{\mbox{$H_X$}\xspace}
\newcommand{\Bx}{\mbox{$B_X$}\xspace}
\newcommand{\Sc}{\mbox{$S_C$}\xspace}
\newcommand{\Hc}{\mbox{$H_C$}\xspace}
\newcommand{\Bc}{\mbox{$B_C$}\xspace}
\newcommand{\PLI}{\mbox{$\Gamma$\Ss{ph}}\xspace}
\newcommand{\lnls}{{log$N$-log$S$}\xspace}
\newcommand{\Deg}{\mbox{$^\circ$}\xspace}
\newcommand{\x}{\mbox{$\times$}}

\newcommand{\Qt}[1]{\mbox{$Q_{#1}$}\xspace}
\newcommand{\Ex}[1]{\mbox{$E_{#1\%}$}\xspace}
\newcommand{\up}[1]{\raisebox{1.5ex}[0pt]{#1}}

\newcommand{\psf}[1]{\mbox{$r_{#1\%}$}\xspace}
\newcommand{\psfp}[1]{\mbox{$r'_{#1\%}$}\xspace}
\newcommand{\psfpp}[1]{\mbox{$r''_{#1\%}$}\xspace}
\newcommand{\psfsq}[1]{\mbox{$r^2_{#1\%}$}\xspace}

\newcommand{\noO}{15\xspace}

\begin{document}

\title{X-ray Processing of ChaMPlane Fields: \\
Methods and Initial Results for Selected Anti-Galactic Center Fields}

\slugcomment{To appear in ApJ}

\author{
JaeSub Hong\altaffilmark{1*},
Maureen van den Berg\altaffilmark{1},
Eric M. Schlegel\altaffilmark{1}, 
Jonathan E. Grindlay\altaffilmark{1}, \\
Xavier Koenig\altaffilmark{1}, 
Silas Laycock\altaffilmark{1}, 
and Ping Zhao\altaffilmark{1}
}
\altaffiltext{*}{Send requests to J. Hong at jaesub@head.cfa.harvard.edu}
\altaffiltext{1}{Harvard-Smithsonian Center for Astrophysics, 
60 Garden St., Cambridge, MA 02138 }

\begin{abstract} 

We describe the X-ray analysis procedure of the on-going \Chandra
Multiwavelength Plane (\champlane) survey and report the initial results
from the analysis of \noO selected anti-Galactic center observations
($90\Deg < l < 270\Deg$).  We describe the X-ray analysis procedures for
ChaMPlane using custom-developed analysis tools appropriate for Galactic
sources but also of general use: optimum photometry in crowded fields
using advanced techniques for overlapping sources, rigorous astrometry and
95\% error circles for combining X-ray images or matching to optical/IR
images, and application of quantile analysis for spectral analysis of
faint sources. We apply these techniques to 15 anti-Galactic center
observations (of 14 distinct fields) in which we have detected 921 X-ray
point sources. We present \lnls distributions and quantile analysis to
show that in the hard band (2 -- 8 keV) active galactic nuclei dominate
the sources.  Complete analysis of all ChaMPlane anti-Galactic center
fields will be given in a subsequent paper, followed by papers on sources
in the Galactic center and Bulge regions.
\end{abstract}


\section{Introduction}

The on-going \Chandra Multiwavelength Plane (\champlane) survey 
is designed to constrain the low-luminosity 
accretion source content and stellar-coronal source luminosity 
functions in the Galaxy \citep{Grindlay03,Grindlay05}.  To achieve this
goal we search for low-luminosity Galactic X-ray sources in \Chandra
archival data and attempt to identify them using follow-up optical and/or
infrared (IR) imaging and spectroscopic observations.

This paper describes the X-ray analysis of the \champlane survey.  To
illustrate the procedures, we describe the initial X-ray results from
the analysis of \noO selected anti-Galactic center (anti-GC) observations
($90\Deg < l < 270\Deg$; Table~\ref{t:list}).  \S\ref{s:ana} describes
the custom analysis tools developed for the project.
\S\ref{s:results} presents initial results including \lnls
distributions of the sources and basic source properties using a new
X-ray spectral classification technique -- quantile analysis \citep{Hong04}.  

In  subsequent papers, we will present the X-ray results of 
all anti-GC fields and the Galactic-Bulge and Center region sources.
The overview of the project is given in \citet{Grindlay05}.
The methodology and initial results of the optical and IR surveys
are presented in \cite{Zhao05}, \cite{Rogel05} and \cite{Laycock05},
respectively.

\begin{table*}
\small
\begin{center}
\caption{15 selected \champlane observations in the anti-GC region (sorted
by Galactic longitude except for the stacked data)\label{t:list}}
\begin{tabular}{r@{\hspace{3mm}}lcrrc@{\hspace{4mm}}r@{}l@{\hspace{5mm}}c@{\hspace{4mm}}c@{}c@{}c@{}c@{}c@{}c@{}c@{}c@{}c@{}cc@{}}
\hline\hline
   		& 		&$l$ 		& $b$ \ \ \ \ 	&Exposure\sS{a} &\nHx\sS{b}  		& \multicolumn{2}{@{}c}{No. of\ \ }		& Obs.~Date	&\multicolumn{10}{c}{CCDs} 		& Aim\sS{d}	 \\
\up{Obs.~ID}	&\up{Target}	&(deg) 		&(deg)\ \  	&(ksec)	\ \ \ 	&($\x10^{22}$cm$^{-2}$) & \multicolumn{2}{@{}c}{Sources\sS{c}} 		& (y-m-d) 	&\multicolumn{10}{c}{used}		& ACIS-\\
\hline
2787 	&PSR J2229+6114	&106.64900 	&2.94848 	& 91.5\ \ \ \	&1.04 / 0.40	& \ \ \ 91& 	&02-03-15	& 0&1&2&    3&.&.&6&    7&.&.&I 	\\
1948 	&3EG J2227+6122	&106.64901 	&2.94965 	& 14.7\ \ \ \ 	&1.02 / 0.40	& \ \ \ 20&  	&01-02-14	& 0&1&2&    3&.&.&6&    7&.&.&I 	\\
755 	&B2224+65 	&108.63800 	&6.84522 	& 47.6\ \ \ \ 	&0.44 / 0.23	& \ \ \ 78&  		&00-10-21	& .&.&2&    3&.&5&6&    7&8&.&S 	\\
2810 	&G116.9+0.2 	&116.94330 	&0.18420 	& 48.8\ \  \ \	&0.48 / 0.28	& \ \ \ 97&  		&02-09-14	& 0&1&2&    3&.&.&6&    7&.&.&I 	\\
2802 	&G127.1+0.5 	&127.11343 	&0.53889 	& 19.2\ \  \ \	&0.89 / 0.31	& \ \ \ 44&  		&02-09-14	& 0&1&2&    3&.&.&6&    7&.&.&I 	\\ \hline
782 	&NGC1569 	&143.68323 	&11.24151 	& 93.3\ \  \ \	&0.42 / 0.11	& \ \ \ 89&  		&00-04-11	& 0&1&2&    3&.&5&.&    7&.&.&S 	\\
650 	&GK Persei 	&150.95713 	&$-$10.10413 	& 90.3\ \  \ \	&0.20 / 0.10	& \ \ \ 83&  		&00-02-10	& .&.&2&    3&.&5&6&    7&8&.&S 	\\
2218 	&3C 129 	&160.42891 	&0.13717 	& 30.2\ \  \ \	&0.63 / 0.24	& \ \ \ 25&  		&00-12-09	& .&.&.&    .&4&5&6&    7&8&9&S 	\\
676 	&GRO J0422+32 	&165.88229 	&$-$11.91286 	& 18.7\ \  \ \	&0.19 / 0.09	& \ \ \ 67&  		&00-12-09	& 0&1&2&    3&.&.&6&    .&8&.&I 	\\
2803 	&G166.0+4.2 	&166.13140 	&4.33778 	& 29.3\ \  \ \	&0.38 / 0.19	& \ \ \ 53&  		&02-01-30	& 0&1&2&    3&.&.&6&    7&.&.&I 	\\ \hline
829 	&3C 123 	&170.58315 	&$-$11.66140 	& 46.3\ \  \ \	&0.59 / 0.12	& \ \ \ 52&  		&00-03-21	& .&.&2&    3&.&5&6&    7&8&.&S 	\\
2796 	&PSR J0538+2817	&179.71974 	&$-$1.68585 	& 19.4\ \  \ \	&0.80 / 0.26	& \ \ \ 31&  		&02-02-07	& .&.&2&    3&.&5&6&    7&8&.&S 	\\
95 	&A0620$-$00 	&209.95777 	&$-$6.54014 	& 41.2\ \  \ \	&0.29 / 0.18	& \ \ \ 61&  		&00-02-29	& .&.&.&    3&.&5&6&    7&8&.&S 	\\
2553 	&Maddalena's Cloud\ \ &216.73098 &$-$2.60034 	& 24.5\ \  \ \	&0.93 / 0.27	& \ \ \ 59&  		&02-02-08	& 0&1&2&    3&.&.&6&    7&.&.&I 	\\
2545 	&M1$-$16 	&226.80033 	&5.62592 	& 48.6\ \  \ \	&0.13 / 0.12	& \ \ \ 71&  		&02-02-11	& .&.&2&    3&.&5&6&    7&8&.&S 	\\ \hline
\sS{e}52787 	&2787 \& 1948 Stacked & 106.64900 & 2.94848  	& 106.2\ \ \ \ 	&1.04 / 0.40	& \ \ \ 90& 	& --		& 0&1&2&    3&.&.&.&    .&.&.&I 	\\ \hline
\end{tabular}
\end{center}
\sS{a}Sum of the good time intervals (GTIs, \S\ref{s:detect}). \\
\sS{b}Two estimates; \citet{Schlegel98} / \citet{Drimmel03}. 
	The total sum along the line of the sight, averaged over source positions. 
	See \S2.2.2 for the differences in the two models.\\
\sS{c}Number of valid (level 1) sources found in the \Bx band. 
	See Tables 2 and 4 for the definition of bands and levels. \\
\sS{d}Aim point detector: I = CCD 3 and S = CCD 7. \\
\sS{e}The combined data (CCD 0, 1, 2, 3) of two stackable observations (1948 and 2787).
	See \S\ref{s:detect} and Table \ref{t:stack}. \\
\end{table*}

\setcounter{footnote}{1}
\section{Data Analysis} \label{s:ana}

The overall data analysis procedures can be grouped into two
stages. First, we search for X-ray point sources using a wavelet detection
algorithm (\wavdetect; \citealt{Mallat98,Freeman02};
\S\ref{s:detect}). Second, we determine various source properties
using simple aperture photometry (\S\ref{s:props}).  To apply the
above algorithms consistently on the large \champlane data set, we
employed a custom X-ray analysis tool (XPIPE) and
have developed post-XPIPE procedures (PXP). Both tools
are primarily based on CIAO tools (version
3.1).\footnote{http://cxc.harvard.edu/ciao.}

XPIPE is an X-ray analysis tool developed for the \Chandra
Multiwavelength Project (\champ) which is a high-latitude
extra-Galactic survey for active galactic nuclei (AGN) (Kim et
al.~2004a and 2004b; K04 hereafter). The scientific goals of the two
projects, \champ and \champlane, are as different as the survey regions
-- Galactic Plane survey versus high-latitude survey.  However, both
projects share similar requirements for the analysis, namely,
searching for faint X-ray point sources in \Chandra ACIS archival
data.  This analysis similarity and the success of XPIPE in the \champ
analysis (K04) led us to adopt XPIPE as our primary tool.  The
detailed description of XPIPE can be found in K04, and in what follows
we only briefly review the basics that are relevant for \champlane.

In addition to XPIPE and in order to meet \champlane-specific analysis
requirements, we have also developed our own analysis tool, PXP, which
is described below.  PXP uses XPIPE outputs and performs analysis that
is optimized for Galactic sources but also of general use.

\subsection{Source detection by {\it wavdetect}} \label{s:detect}

For each observation, we start the analysis by using XPIPE on the
level 2 data products generated by the \Chandra X-ray Center (CXC)
standard data processing. First, XPIPE removes the residual artifacts
that may be present after the CXC standard data processing.  XPIPE
then selects events in the good time intervals (GTIs)\footnote{These are
different from the GTIs set by the CXC processing (K04).} which are
selected to have fluctuations $ < 3 \sigma$ from the mean background rate.

\begin{table}[b]
\small
\begin{center}
\caption{Energy bands for the \champlane X-ray analysis \label{t:eband}}
\begin{tabular}{lr@{\hspace{2mm}}c@{\hspace{2mm}}c}
\hline\hline
Type		&& Name		&	Range		\\
\hline
\it XPIPE \wavdetect band 	&		&		\\
& Soft 	& \Sx		& 	0.3 -- 2.5 keV		\\
& Hard 	& \Hx		& 	2.5 -- 8.0 keV		\\
& Broad	& \Bx		& 	0.3 -- 8.0 keV		\\
\hline
\it Conventional band \ \ \ \ \ \ \ 	&		&	\\
& Soft 	& \Sc		& 	0.5 -- 2.0 keV		\\
& Hard 	& \Hc		& 	2.0 -- 8.0 keV		\\
& Broad	& \Bc		& 	0.5 -- 8.0 keV		\\
\hline
\end{tabular}\\
\end{center} 
\end{table}

For source detection, XPIPE uses \wavdetect with a significance
threshold of $10^{-6}$ (about 1 possibly spurious detection per CCD, see K04
for the detailed analysis of source detection efficiency with
\wavdetect in XPIPE) and a scale parameter varying in seven steps
between 1 and 64 pixels to cover a wide range of source sizes. XPIPE
applies \wavdetect in three separate energy bands, \Sx, \Hx and \Bx (Table
\ref{t:eband}), using the exposure map generated at 1.5 keV.  Note
that the \wavdetect routine in CIAO version 3.1 includes source
position refinement procedures that had to be applied separately in
the old version of XPIPE based on CIAO version 2.3 (K04).

When multiple observations of a similar region of the sky are available,
it is possible to combine the data sets to look for fainter sources and
to study variability.  Particularly near or at the GC region, many
stackable observations are available \citep{Grindlay05}. We have devised
PXP so that it can employ \wavdetect on stacked images.  We consider
multiple observations stackable if their aimpoints are within 1$'$
of each other and the aimpoint detectors are the same type (ACIS-I
or ACIS-S).  Among the stackable observations, we designate the one with
the longest exposure to be the base observation of the stacked image.
For the analysis convenience, we assign a separate Obs.~ID for the stacked
data ($xyyyy$, where $x$ is $> 5$ and $yyyy$ is the base Obs.~ID) and
we assign the CCD ID of the stacked data to be 3 for ACIS-I and 7 for
ACIS-S observations.  For stacking, we only use the data of CCD 0, 1,
2, 3 for ACIS-I and the data of CCD 7 for ACIS-S observations.
For the aimpoint, detector response function, and other required
parameters of the stacked data, we employ the same of the base observation.
In the anti-GC fields (Table \ref{t:list}), Obs.~ID 52787 is the stacked
data of two observations (1948 and 2787) and Obs.~ID 2787 is the base
observation of the two.

Before we stack the data, we apply astrometric
corrections (see also \S\ref{s:sim_err}). First, we calculate the
correction for the aspect offset of each
observation.\footnote{http://cxc.harvard.edu/ciao/threads/arcsec\_correction} 
Second, we calculate the boresight offset of each observation relative to
the base observation using the aspect-corrected position of the {\it
wavdetect}ed sources.  The final boresight offset of each observation is
derived by an iterative procedure using a relatively more reliable subset
of matching source pairs.
The iterative technique is identical to the
boresight correction procedure employed between X-ray and optical data
for the optical \champlane survey \citep{Zhao05} except that the
procedure is applied between X-ray and X-ray data using the estimated
95\% X-ray positional errors (\S\ref{s:sim_err}).  

For the purpose of stacking the X-ray data pointed to a similar region
of the sky (i.e. large overlap in the field of view), direct comparison
of the boresight of an X-ray observation to another X-ray observation
is more efficient than comparing it to observations (or catalogs) in
another wavelength because the number of matching pairs
found in the former case is likely larger than that in the latter and a
large number of matching pairs usually provide a reliable estimation of
the relative boresight offset.  In the case of Obs.~ID 1948 and 2787,
we have 18 matching pairs out of possible 20. According to source
net count ($> 10$), off-axis angle ($< 8'$) and level ($=1$; see
\S\ref{s:stat}), 12 of them are selected for calculating the final
boresight offset at the end of the iteration procedure
(Table~\ref{t:stack} and see \S\ref{s:sim_err}).

Once the aspect and boresight offsets are calculated, PXP reprojects
all the stackable data (XPIPE-screened event files) onto a common
projection point (the aimpoint of the base observation) accordingly,
and applies \wavdetect on the stacked image in the \Bx band using the
total exposure map (at 1.5 keV) for the stacked data.  In the case of
Obs.~ID 52787 with respect to two Obs.~IDs 1948 and 2787, we detect 8 new
sources and miss 5 sources by stacking (Table~\ref{t:stack}).

\begin{table}[t]
\small
\begin{center}
\caption{Source detection in the stacked image (Obs.~ID 52787) 
of two observations (1948 and 2787)\label{t:stack}}
\begin{tabular}{rr@{\hspace{2mm}}cccc@{\hspace{2mm}}rr}
\hline\hline
Obs.\	& Exp.	 & \multicolumn{2}{c}{Offset (R.A., Dec.)}	&& \sS{a}No. of 	&\multicolumn{2}{c}{Common} 	\\ \hhline{~~--~~--}
ID\ \ 	& (ksec) & Aspect		& Boresight\sS{c}	&& sources	&1948		&2787		\\ \hline
1948	& 14.7	 &$-1.82''$, 0.62$''$	&$-0.43''$, $-0.05''$	&& 20		&--		&\sS{d}18	\\
2787	& 91.5	 &$-0.40''$, 0.03$''$	&--			&& 85		&\sS{d}18	&--		\\ \hline
52787	& 106.2	 &--			&--			&& 90		&20		&80		\\
\hline
\end{tabular}\\
\end{center} 
\sS{a}Number of valid (level 1) sources in CCD 0, 1, 2, and 3. Note that
 Obs.~2787 has 6 sources in CCD 6 and 7 (see Table~\ref{t:list}).\\
\sS{b}Number of common sources found in both data set. This is
	determined by the X-ray positional error (\S\ref{s:sim_err}).\\
\sS{c}Relative to the base Obs.~ID 2787 after correcting aspect offsets. 
	Among 18 matching pairs, 12 pairs are selected 
	for calculating the final boresight offset. \\
\sS{d}The other two sources of Obs.ID 1948 are located in the chip gap of Obs.~ID 2787. 
\end{table}

\subsection{Source properties by aperture photometry} \label{s:props}

Subsequently, PXP employs the source detection results of XPIPE
(or PXP for stacked data) in the \Bx band and extracts basic source
properties via aperture photometry on XPIPE-screened event files using
the ``conventional'' bands -- \Sc,
\Hc and \Bc -- shown in Table \ref{t:eband}.  Note that while XPIPE does
detection in three bands, we only consider detections in the \Bx band.
XPIPE also performs an aperture photometry using all 6 bands in Table
\ref{t:eband}, but we implement a separate aperture photometry in PXP
(modelled after the one in XPIPE) to be effective in crowded fields
(e.g.~GC) and to provide more versatile outputs that are particularly
useful for describing the diversity of source populations found in the
Galactic Plane. The aperture photometry in PXP is also designed to be
compatible with the stacked data.

Note also that \wavdetect provides basic source properties (e.g., net counts).
However, we employ \wavdetect with a 39\% inclusion radius of the point spread
function (PSF; \psf{39}: 39\% of source photons lie within the
circle\footnote{The \psf{x} values are calculated at 1.5 keV from the \Chandra
calibration data {\it
psfsize\_20010416.fits} (http://cxc.harvard.edu/cal/Acis/).}),
which is recommanded by \cite{Freeman02} for the simple algorithm of
source characterization in \wavdetect.  Therefore, source properties
reported by \wavdetect may not be accurate due to the large missing
fraction of the source region particularly when the source is near
large regions of diffuse emission.

\begin{table*}
\small
\begin{center}
\caption{Aperture photometry in \champlane analysis \label{t:ap}}
\begin{tabular}{c@{\hspace{5mm}}llc@{\hspace{5mm}}cc@{\hspace{5mm}}cc@{\hspace{5mm}}c}
\hline\hline
Source 		&						&		&& Core Radius			&&  Refined 		&& Background \\ 
Region Overlap	&\multicolumn{2}{c}{\up{Condition}}				&& ($r_c$)			&&  Source Region	&& Region				\\ \hline
No  		&\multicolumn{2}{@{}l}{$\Delta\ge \psf{95} + \psfp{95}$} 	&& $\psf{95} $			&& $r \le r_c=\psf{95}$	&&					\\ \hhline{-------~}
  		& $\Delta\ge 1.5\,\psfp{95}$, 	& $\Delta<\psf{95} + \psfp{95}$	&&  				&& $r \le r_c$ and 	&& $2\,\psf{95} < r < 5\,\psf{95} $ and\\ \hhline{~--~~~}
Yes  		& 				& $\Delta\ge\psf{68}+\psfp{95}$	&& \up{$\Delta - \psfp{95}$}	&&  pie sector in	&& $r'' > \psfpp{95}$ for all neighbors	\\ \hhline{~~---~~}
		& \up{$\Delta< 1.5\,\psfp{95}$,}& $\Delta<\psf{68}+\psfp{95}$	&& $\Delta/3$			&& $r_c<r\le \psf{95}$	&&\\ \hline
\end{tabular}
\end{center}
Notes.~--- $2.5''\le \psf{95}\le 40''$. $\Delta$ is the distance between the source and the nearest neighbor,
and \psfp{95} is the 95\% PSF radius of the nearest neighbor. At a given
position in the sky, $r$ is the distance from the source
and $r''$ is the distance from neighbors with the
95\% PSF \psfpp{95}.  Note that when there is an overlap, $\psfp{95}
\simeq \psf{95}$ because of the relatively small change of the PSF
size compared to the change in source position ($<2\%$).
When overlapping is severe ($r_c < \psf{39}$ and $1/R' < 0.3 $),
we return to a simple aperture photemetry (flag=142). See Eq.~(\ref{e:sec}) for $R'$.
The PSF radii are calculated at 1.5 keV.
\end{table*}

Table \ref{t:ap} summarizes the key parameters of the aperture
photometry in PXP.  We define the basic source region using a circle
around the source position ($\le \psf{95}$); the background region is
defined by an annulus ($ 2\,\psf{95} < r < 5\,\psf{95}$).  
For practical purposes, we limit the range of \psf{95} to $2.5''\le
\psf{95}\le 40''$.\footnote{The actual minimum value of \psf{95} is $\sim 2''$.
In order to compensate for a possible positional error, we add
0.5$''$, which is the expected positional error for $\sim$ 20 count
sources at $\lesssim 3'$ away from the aimpoint.  See \S\ref{s:sim_err}.}
  If the background annulus overlaps with
neighboring source regions we exclude the neighboring source regions
from the background region (Table~\ref{t:ap}).

\begin{figure}[b] \begin{center} 
\epsscale{.575}
\plotone{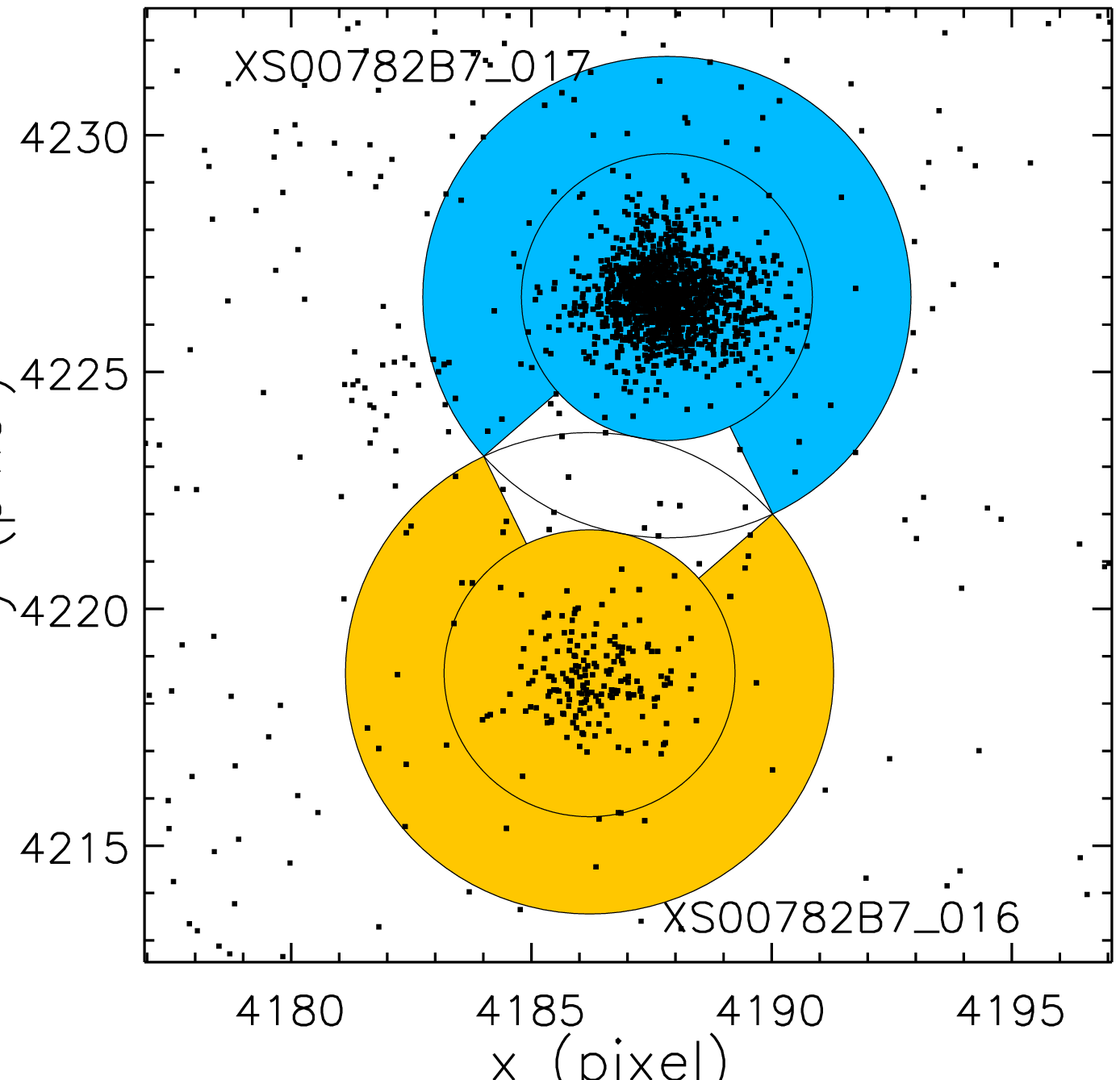} 
\plotone{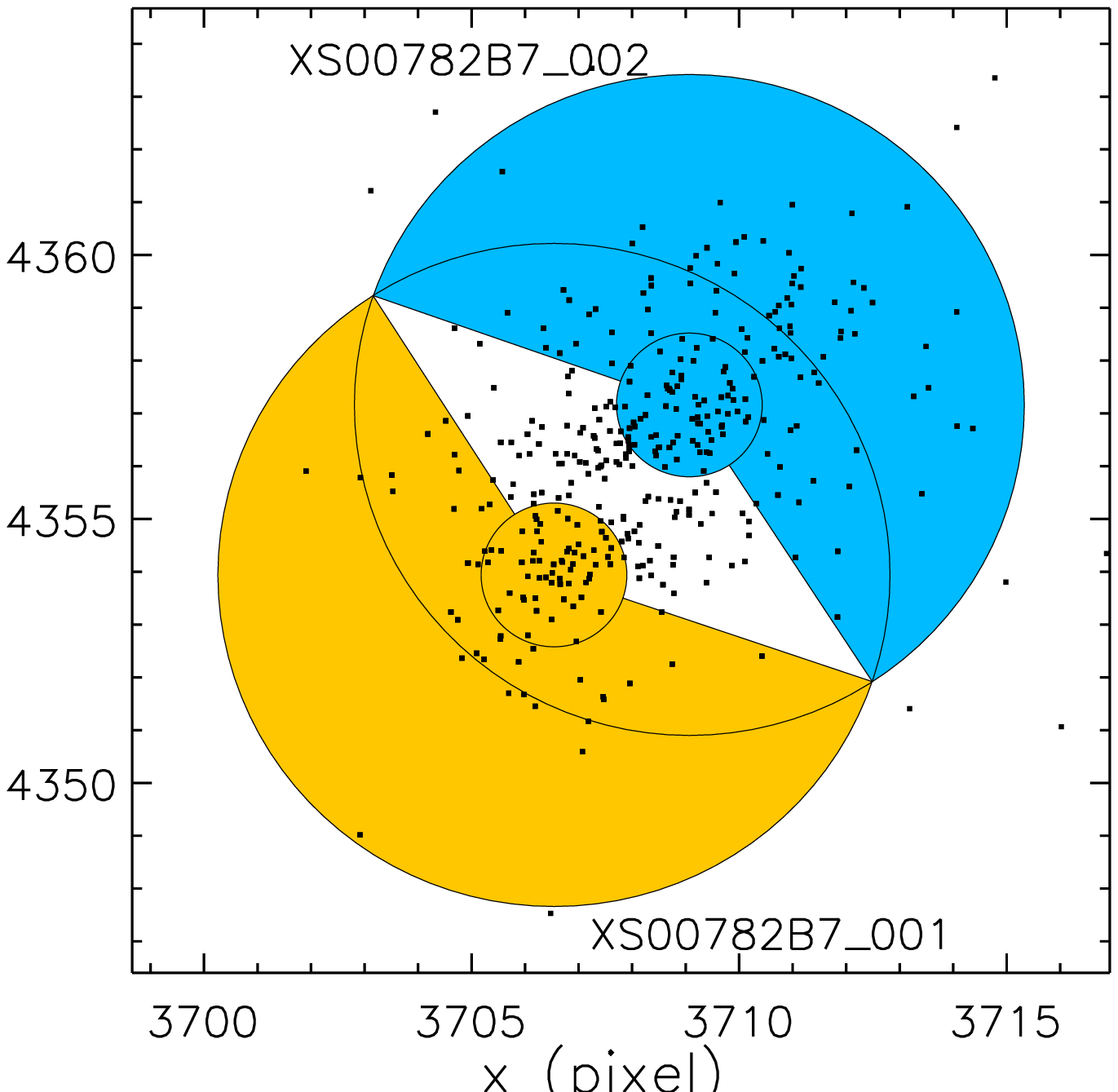} 
\end{center} 
\caption{Examples of overlapping source regions; relatively small
overlap (left) and large overlap (right).
The refined source region (yellow shade for one and blue for
the other source) consists of a core and the uncommon pie sector.
In the case of XS00782B7\_001 and XS00782B\_002 (right panel),
this refinement in aperture photometry reveals 
these two sources have significantly different spectral types, which are 
otherwise indistinguishable (see \S\ref{s:qt}). 
See the electronic ApJ
for the color version of the figure }
\label{f:ap} \end{figure}

Source regions that overlap are handled in the following manner.
We divide the source region into a core (circle) and a shell (annulus).
We refine the source region to be the sum of the core and a pie sector
of the shell that excludes the common sector with the neighbor's source
region.  The core size is determined empirically to include as many source
photons as possible, while minimizing contamination from neighbors. If
the source region overlaps with multiple neighbors, the core radius is
determined by the nearest neighbor, and the pie sector excludes all the
common sectors with the neighbors' source regions.

Fig.~\ref{f:ap} shows examples of overlapping source regions.  The shaded
region indicates the refined source region: the core and the uncommon
pie sector (yellow shade for one and blue shade for the other source).
In the left panel of Fig.~\ref{f:ap} where the overlap is relatively
small, the core region does not overlap with the neighbor's source region.
In the right panel, the core radius is set to be one-third of the
distance between two sources due to relatively large overlap.
Simple aperture photometry using the source region with $r\le\psf{95}$
produces almost identical spectral properties for the two sources
in the right panel; XS00782B7\_001 and XS00782B7\_002.\footnote{Source
ID follows the \champ converntion: XS\{Obs.~ID\}\{energy band\}\{CCD
ID\}\_\{Source No.\} and the \Bx band is the {\it B} band in \champ.}
Aperture photometry employing the refined source region reveals a
significant difference in the spectral types of these two sources
(\S\ref{s:qt} and Fig.~\ref{f:qccd}).  

Note that the above correction for overlapping source region is modified
from the original correction implemented in XPIPE (K04; Kim 2005) in
order to allow relatively high source counts and easy quantile analysis
(\S\ref{s:qt}).

XPIPE sets various flags on each source detected by \wavdetect
depending on source properties such as proximity to a chip
boundary. These flags allow us to determine the validity of each
source.  A subset of the flags are set manually by the V\&V step (K04)
in order to remove residual artificial sources such as those stemming
from frame transfer streaks during read-outs.  PXP inherits the XPIPE
V\&V-based source flag information and sets additional flags relevant
to the aperture photometry routine in PXP.

Table~\ref{t:flag} lists the flags and their definitions.  For example,
when the source region overlaps with neighbors' source regions, we flag
the source (141).  If the core for the refined source region is very
small ($r_c < \psf{39}$) and the area of the uncommon pie sector is
less than 30\% of the shell area, we return to the simple aperture
photometry because the refined aperture photometry is also subject to
a large uncertainty and we set another flag (142) on the source for
further analysis.

For the initial analysis, we extract three groups of source properties
from the aperture photometry: source count and rates, flux, and quantiles.

\begin{table*}
\small
\begin{center}
\caption{Source Flags used in \champlane Analysis\label{t:flag}}
\begin{tabular}{l@{\hspace{5mm}}l@{\hspace{5mm}}l}
\hline\hline
Type			& Flag		&	Definition		\\
\hline
\multirow{6}{25mm}{\it False sources by V\&V\sS{a}}
			& 111		& False source by hot pixels or by bad bias values \\
			& 112		& False source by bad columns \\
			& 113		& False source due to readout streaks by a very strong source \\
			& 114		& False source by the FEP 0/3 problem\sS{b} \\
			& 115		& Double source detected by the PSF effect\sS{c} \\
			& 121		& Other spurious sources \\
	
\hline
\multirow{5}{25mm}{\it Valid sources but source properties may be subject to a large uncertainty\sS{d}}
			& 141		& Source region overlaps with neighbors' source regions \\
			& 142		& No refinement on source region due to severe source region overlap \\
			& 143		& Background region overlaps with neighbors' source regions	\\
			& 146\sS{e}	& Source at the chip boundary \\
			& 147		& Pile-up candidate ($\ge$ 100 cts/ksec)	\\		
			& 148		& Uncertain source position by flag = 115\sS{c} \\
\hline
\multirow{5}{25mm}{\it Other cases}
			& 151		& Source is extended \\
			& 153		& Target of the observation	\\
			& 154		& Sources in the target region (e.g. point sources in a target cluster) \\
			& 157\sS{f}	& The same source is found in other observations	\\
			& 158\sS{f}	& The same source is found in the stacked image	\\
\hline
\end{tabular}\\
\end{center}
Notes.~--- XPIPE assigns a very similar set of flags (see Table 3 in K04),
	but we redefine a full set of flags for PXP because many flags
	depend on the details of the aperture photometry. 
	To avoid confusion with the XPIPE flags, the PXP flag IDs $\ge 100$ and
	the XPIPE flag IDs $< 100$. \\
\sS{a}Provided by the XPIPE Validate and Verify (V\&V) record. \\
\sS{b}http://cxc.harvard.edu/ciao/caveats. \\
\sS{c}See Fig.~7 in K04. \\
\sS{d}These flags are set automatically. The 95\% PSF size for source
	region is based on {\it psfsize\_20010416.fits} (\S\ref{s:props}). \\
\sS{e}When the source region contains any pixel that has 10\% or less of the maximum exposure
	value of the CCD. \\
\sS{f}These flags are set automatically, based on Eq.~(\ref{e:cross}). \\
\end{table*}

\subsubsection{Net counts and net count rate}

For a given band, the number of net counts ($N\Ss{net}$) of a
source without source region overlap is derived from the number of
counts ($N\Ss{src}$)
in the source region (with area $A_S = \pi \psfsq{95}$), the number of
counts ($N\Ss{bkg}$) in the background region (with area $A_B$),
and the relative ratio ($R$) of the two areas scaled by the
exposure map ($e_p$) values of the regions.
\begin{eqnarray}
	N\Ss{net} = N\Ss{src} - R\  N\Ss{bkg},\  \mbox{\ and\ \ } 
	R = \frac{A_S\  \overline{e_p}|_{p \in A_S}}{\sum_{p \in A_B} e_p},
	\label{e:net}
\end{eqnarray}
where $\overline{e_p}|_{p \in A}$ is the mean value of $e_p$ in $A$.
Note that ideally $A \ \overline{e_p}|_{p \in A} = \sum_{p \in
A} e_p$, but when $A$ is small (often true for source regions), the
former produces more reliable estimates of the exposure sum because
exposure maps are usually pixellated.

In the case of source region overlap, we assume azimuthal symmetry of the
PSF and we derive $N\Ss{src}$ from the number of counts ($N\Ss{core}$)
in the core and the number of counts ($N\Ss{pie}$) in the uncommon pie sector
(with area $A_P$),
\begin{eqnarray}
	N\Ss{src} = N\Ss{core} + R'\  N\Ss{pie},\  \mbox{\ and\ \ } 
	R' = \frac{A'_{P}+A_P}{A_P}, \label{e:sec}
\end{eqnarray} where $A'_{P}$ is the area of the common pie sector
and we use Eq.~(\ref{e:net}) for $N\Ss{net}$.

The net count rate ($r\Ss{net}$) is defined by the ratio of the number of
net counts to the exposure time ($T$) and it is scaled by the ratio of
the mean value of exposure map within the source region to the value
at the reference point, $\mbox{max}(e_p)$,
\begin{eqnarray}
	r\Ss{net} = \frac{N\Ss{net}}{T} \frac
		{\mbox{max}(e_p)}
		{\overline{e_p}|_{p \in A_S}}.
	\label{e:rate}
\end{eqnarray} 

The max($e_p$) is the maximum of the exposure map value of the chip.
This exposure map scaling is implemented for easy conversion from count
rate to flux regardless of source position on the chip.

PXP calculates the number of net counts, the net count rate and
their errors for each source in the \Sc, \Hc, \Bc, and \Bx bands
(Table~\ref{t:eband}).

\subsubsection{Flux} \label{s:flux}

To estimate the X-ray flux from a source, PXP uses
{\sherpa}\footnote{http://cxc.harvard.edu/sherpa/threads/index.html}
to calculate the count rate to flux conversion factor. To do so, we
need an accurate detector response function, a column density (\nH)
to the source, and a reliable spectral model for X-ray emission.

The detector response function varies with source position on the chip.
However, we simply use the redistribution matrix file (RMF) and the
auxiliary response file (ARF) calculated at the reference point (exposure
maximum) for each CCD because spatial variation of RMFs and ARFs is
relatively small (usually $<5\%$).\footnote{The CIAO version 3.2 can
generate the spatially variant RMF and ARFs. The analysis
presented in this paper is performed under the CIAO version 3.1.}
Note that the ARFs generated by {\it mkarf} account
for the known temporal degradation of the low energy efficiency of ACIS
detectors.\footnote{Implemented in the {\it mkarf} of the CIAO tools
(version 3.0 or higher). See http://cxc.harvard.edu/cal/Acis/.}
For stacked data, we simply use the RMFs and ARFs of the base observation
for the initial estimation.

For the spectral model or column density, as we do not have an {\it a
priori} prescription or estimate for most detected sources, we use
several simple models to estimate the source flux, viz. power-law
models with $\PLI = $1.7, 1.4 and 1.2 
likely to be appropriate for accretion powered sources such as
background AGN and X-ray binaries. 
We also use Raymond-Smith, MeKaL, 
thermal Bremsstrahlung, and Black Body -- all with $kT=1.0$ keV -- 
for thermal sources (e.g., coronal sources).\footnote{Corresponding model
names in \sherpa
are {\it xspowerlaw, xsraymond, xsmekal, xsbremss} and {\it  xsbbody} 
with {\it xswabs} for the column density.}

For the column density \nH, we employ two models; \citet{Schlegel98}
and \citet{Drimmel03}.  In general, \citet{Schlegel98} provide
realistic estimates of total Galactic extinction, but their estimation
near the Galactic Plane ($|b| < 5\Deg$) is not as reliable as in
high-latitude fields.  The model may overestimate the extinction due
to incomplete correction for point source contributions in the Plane.
\citet{Drimmel03} provide a 3-D model of extinction in the Galaxy,
which is useful for Galactic sources with known
distances.\footnote{There is a distance parameter in the model by
\citet{Drimmel03}. The model has only Galactic components, so that its
estimate does not change for distances greater than $\sim$ 20 or 30
kpc.}  However, their model may underestimate the total column density
since it does not include contributions from all known molecular
clouds.  

As shown in Table 1, the \nH estimates by \citet{Schlegel98} for
the selected anti-GC fields are indeed larger (by $\sim 10 - 200 \%$
depending on the fields) than the estimates by
\citet{Drimmel03}. According to \citet{Drimmel03}, their model 
is normalized so that the two models agree well for high
latitude fields, but as indicated by large disagreements, both models
suffer relatively large uncertainties at low latitude fields.

In addition to intrinsic uncertainties in the models, we do not know the
distance of most of \champlane sources.  Therefore, PXP calculates the
\nH value at each source position for both models, assuming the total
column density along the line of the sight.
The angular resolutions of the models are $\sim 6'$ for
\citet{Schlegel98} and $\sim 20'$ for
\citet{Drimmel03}, and $\nH$ is then interpolated on this grid to the
source position.  We use $\nH/E(B-V) = 5.8 \times 10^{21}$
cm\sS{-2}/mag \citep{Bohlin78} to convert \citet{Schlegel98} values of
E(B-V) and $\nH/A(V) =1.79 \times 10^{21} $ cm\sS{-2}/mag
\citep{Predehl95} with the ``rescaled'' option to account for the
small angular scale dust structure in using the \cite{Drimmel03} maps.

For each source, by default, PXP estimates the X-ray flux in the \Sc,
\Hc, \Bc and \Bx band using all seven spectral models and two extinction
models. PXP is designed to allow easy addition of new spectral or
extinction models and it can also assign a separate spectral and
extinction model unique to each source.  In future, for sources
with information on spectral type, distance or column density, the
X-ray flux will be revised accordingly.  For example, one can estimate
the column density to a source from the quantile analysis of the X-ray
spectrum if one can assign a reliable spectral model to the source
(see \S\ref{s:qccd}).

\subsubsection{Quantiles} \label{s:qt}

The survey strategy ultimately relies on the optical or IR
spectroscopy to identify the nature of low luminosity X-ray sources
discovered from the \Chandra archival data
\citep{Grindlay05}. However, even at the R $\sim$ 24.5 optical ChaMPlane
limits (Zhao et al 2005), we do not find counterparts for about half the
ChaMPlane X-ray sources (e.g.~Table~\ref{t:lnls}).  In these cases,
we rely on the X-ray data for a clue to the source nature.  Since the
majority of the sources possess only $\sim$10--20 counts it is important
to use a technique that is sensitive to low-count statistics and with
minimum count-dependent bias.

In the \champ project, source classification by the X-ray data relies
on the X-ray hardness ratio and X-ray colors (K04). These quantities
are based on a particular choice of three sub-energy bands (0.3--0.9,
0.9--2.5, and 2.5--8.0 keV).  In \champlane, it became quickly clear that
this band choice is not optimal for many highly-absorbed Galactic sources
or very soft coronal sources due to count-dependent selection
effects intrinsic to this particular choice of bands \citep{Hong04}. In
fact, for X-ray hardness ratio or colors, there is no single set of
energy bands that can effectively describe the diverse classes of X-ray
sources found in Galactic Plane fields.

Therefore, for \champlane, we employ a new spectral classification
technique, quantile analysis, to acquire more versatile measures of
X-ray characteristics of sources. First introduced by \citet{Hong04},
the quantile analysis employs various quantiles (median, terciles,
quartiles, etc) of spectral distributions to reveal
and classify various spectral features and shapes.  \citet{Hong04}
illustrate the technique using a quantile-based color-color diagram
(QCCD), where the median of the distribution is used to indicate the
overall hardness and the quartile ratio is used to classify the
general shape (concave-up and -down) of the spectrum.

The quantile analysis does not require sub-division (binning) of the
energy range, and it takes full advantage of energy resolution of the
instruments. Therefore, quantile analysis is free of any selection effects
inherent in the conventional approaches, and source classification by
QCCDs is uniformly more sensitive than that by conventional X-ray
color-color diagrams \citep{Hong04}.

For a given source, PXP feeds the lists of photon energies in the source
and background regions together with the weighted ratio of the two
regions ($R$) into the 
routine {\it quantile.pl} 
\citep{Hong04}.\footnote{Quantile analysis routines are available at 
http://hea-www.harvard.edu/ChaMPlane/quantile/.}  
In the case of source region overlap, 
we use events in the refined source region (the core plus the uncommon pie
sector) and the area ratio $R$ is given by
\begin{eqnarray} 
	R = \frac{(A_C+A_P)\  \overline{e_p}|_{p \in A_S}}{\sum_{p \in A_B} e_p},
\end{eqnarray} where $A_C$ is the area of the core and $A_P$ the area
of the uncommon pie sector. 
For example, without refining the 
source region, the two sources in
the right panel of Fig.~\ref{f:ap} show almost identical spectral
properties due to
source photon mixing; the median energy \Ex{50} =
3.96(11) keV for XS00782B7\_001 and \Ex{50} = 3.96(09) keV for 
XS00782B7\_002. However, the above-refined 
aperture photometry reveals
their spectral type is indeed quite different; \Ex{50} = 4.29(15) keV for 
XS00782B7\_001 and \Ex{50} = 3.73(12) keV for XS00782B7\_002 [diamonds
in Fig.~\ref{f:qccd} (c)]

For each source, PXP generates five quantiles, median ($m=\Qt{50}$),
terciles (\Qt{33} and \Qt{67}), quartiles (\Qt{25} and \Qt{75}),
and their errors in two broad energy bands, \Bx and \Bc.  Examples of
quantile analysis on the anti-GC fields are provided in section \ref{s:qccd}.

\subsection{Positional uncertainty and cross-correlating X-ray sources}
\label{s:sim_err}

The \wavdetect routine provides the sky position and positional errors
of each source, but \wavdetect underestimates the positional errors
because the inputs to \wavdetect are not ideal.  For example,
we use symmetric PSFs for \wavdetect but the PSFs become asymmetric at
large off-axis angles. In addition, the exposure map is calculated at 1.5
keV for practical purposes (no information about the source spectrum
is available in advance; the effective area as a function of energy
peaks at 1.5 keV) but 
no 
\champlane source is expected to
exhibit purely 1.5 keV emission. For optical or IR identification of
X-ray sources, we need a more reliable estimate for the uncertainty of
{\it wavdetect}ed source positions.  To do so, we have performed
SAOSAC\footnote{http://cxc.harvard.edu/chart/.} \&
MARX\footnote{http://space.mit.edu/CXC/MARX/.} simulations.

We generate a series of simulated sources evenly spaced in radial
($1'$ spacing out to $\sim 10'$) and azimuthal directions (15\Deg) from the aim
point. We simulate the \Chandra High Resolution Mirror Assembly (HRMA)
PSF using SAOSAC and simulate ACIS-I detection using MARX.  The source
photons are sampled from a power law model with $\PLI=1.5$ and $\nH =
3 \times 10^{20} $ cm\sS{-2} and the background photons from the
Markevitch `period B' ACIS-I background dataset ({\it
acisi\_B\_i0123\_bg\_evt\_230301.fits}).\footnote{
http://hea-www.harvard.edu/$\sim$maxim/axaf/acisbg/. }  The total
source photons for a given source ranges from 5 to 1000 counts, and
the background is integrated over 10 ksec (for $< 15$ count sources)
or 20 ksec (for $\ge 15 $ count sources).

\begin{figure} \begin{center} 
\epsscale{1.00}
\plotone{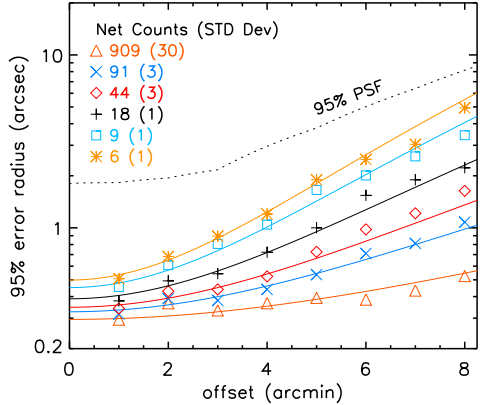} 
\end{center} 
\caption{95\% inclusion radius ($P\Ss{err}$) for uncertainties of {\it
wavdetect}ed source positions by SAOSAC \& MARX simulations; the data
points are simulated results for a given number of detected net counts reported by
\wavdetect, and the solid lines are the visually
chosen empirical formula, Eq.~(\ref{e:sim_err}). The dotted line represents 95\% PSF
radius ($r\Ss{95\%}$, \S 2.3), used for the upper limit of $P\Ss{err}$.}
\label{f:sim_err} \end{figure}

We apply \wavdetect to the simulated data, using a simplified
version of XPIPE, and compare the {\it wavdetect}ed source
positions with the true positions. Fig.~\ref{f:sim_err} shows the size of such
calculated 95\% error circles (95\% of the true source positions lie
within the circle) as a function of the offset from
the aimpoint ($D\Ss{offset}$ in arcmin) and the number of net
counts ($c\Ss{n}$).\footnote{This net count ($c\Ss{n}$) is
the number reported by \wavdetect and it is not the same as
$N\Ss{net}$ which is calculated by the aperture photometry.}  Based on
this result, we generate the empirical
formula for 95\% error circle $P\Ss{err}$ of the {\it wavdetect}ed
source position.
\begin{eqnarray}
P\Ss{err} & &=  0.25'' 
	+ \frac{0.1''}{ \log_{10}(c\Ss{n}+1) } 
	\left[1 + \frac{1}{ \log_{10}(c\Ss{n}+1) } \right] \nonumber \\
	+&& 0.03'' \left[ \frac{D\Ss{offset}}{\log_{10}(c\Ss{n}+2)} \right]^2
	+0.0006'' \left[ \frac{D\Ss{offset}}{\log_{10}(c\Ss{n}+3)} \right]^4.
	\label{e:sim_err}
\end{eqnarray}\\

The simulations are performed for ACIS-I detectors only and we employ
the formula for both ACIS-I and ACIS-S observations.
For very small count sources ($c\Ss{n} \ll 5$), the error estimate
made by the above formula can be unreasonably large, so we limit 
the positional error as $P\Ss{err} \le r\Ss{95\%}$ because
$P\Ss{err} \approx r\Ss{95\%}$ for 
a one count source with no background.

For matching with optical or IR sources, we use the $P\Ss{err}$ 
along with other relevant information such as boresight offset, optical
error circle, etc \citep{Laycock05,Zhao05}.
For the X-ray data base, PXP generates an index table that
cross-correlates potentially identical sources that appear in multiple observations
of the same region.  First, we correct the X-ray source positions
using the known aspect offset for each 
observation.\sS{4}
Next, we consider two sources to be potentially identical (and also set flag=157),
when the distance ($\Delta\Ss{12}$) between the sources is less than
the quadratic sum of the error radius of both sources, i.e.
\begin{eqnarray}
	\Delta\Ss{12} \le (P^2_{\mbox{\scriptsize err1}} +P^2_{\mbox{\scriptsize err2}} + 0.7^2 \x 2 )^{0.5},
	\label{e:cross}
\end{eqnarray}
where the constant term is added to account for the accuracy of the absolute
astrometry (0.7$''$ for 95\% error for each observation) after the
aspect offset
correction\footnote{http://cxc.harvard.edu/cal/ASPECT/celmon/; assume 0.6$''$ for 90\% error.}. 
Note Eq.~(\ref{e:cross}) is used only for cross-correlating 
(or boresighting) X-ray to X-ray observations; for matching with optical or IR
sources, see \citep{Laycock05, Zhao05}.

\section{Initial results for the anti-GC fields} \label{s:results}

In the following, we consider the results from the unstacked data.

\subsection{Source statistics} \label{s:stat}
Table~\ref{t:st} summarizes the initial X-ray results of the
\champlane survey on the \noO selected (see Table 1) 
anti-GC observations.  The table shows the
summary of source counts at each level of the analysis.  The levels are
devised to select valid sources and 
to increasingly (with level) restrict systematic errors. 

The \wavdetect routine detected 1028 sources in the \noO selected anti-GC
observations (level 0).  After careful selection of sources by the flag
information and V\&V, we found 921 reliable sources (level 1) in these
regions.  To interpret the results without concern for anomalies arising
far from the aim point, we further limit the source list.  In level
2, we use only data from CCDs 0, 1, 2, and 3 for ACIS-I observations,
and CCD 2, 3, 6, and 7 for ACIS-S observations.  In level 3, we select
sources within 400$''$ of the aim point for ACIS-I observations, and
all of CCD 7 for ACIS-S observations.  Note that level 3 uses only
front-illuminated (FI) CCDs for ACIS-I and only one back-illuminated
(BI) CCD for ACIS-S observations.

The last three columns in Table \ref{t:st} list the number of sources
that need special attention for further analysis: the same sources found
in multiple observations of the same field (flag=157; i.e., the first 2
observations listed in Table 1), sources that may be piled-up (flag=147),
and sources with their source region overlapping with neighbors' source
region (flag=141).

\begin{table*}
\small
\begin{center}
\caption{Summary of Source Numbers for \noO selected anti-GC observations (7 ACIS-I,
8 ACIS-S Observations.) \label{t:st}}
\begin{tabular}{rl@{\hspace{2mm}}lrrrcrrr}
\hline\hline
		&				& 						& \multicolumn{3}{c}{Number of Sources}		&& \multicolumn{3}{c}{Special\sS{a}}	\\ \hhline{~~~---~---}
		&\up{Stage}			& \up{Selection Rules}				& ACIS-I 	& ACIS-S 	& Total \ \	&& multiple	& pile-up	& overlap	\\ \hline
level 0		&\wavdetect			& Everything					& {448\ }	& {580\ }	& 1028 	\ \	&& 18\x2 	& 3 		& 28		\\ \hline
level 1		&Select valid sources\sS{b}	& Flag $\ne$  11x, 12x, nor 146			& {431\ }	& {490\ }	& {921\ \ }	&& 18\x2		& 3		& 20		\\ \hline
 		& 				& CCD 0, 1, 2, 3 for ACIS-I Obs.,		&		&		& 							\\
\up{level 2}	& \up{CCD selection}		& CCD 2, 3, 6, 7 for ACIS-S Obs.		& \up{409\ }	& \up{445\ }	& \up{854\ \ }	&& \up{18\x2}	& \up{2} & \up{16} 	\\ \hhline{~----}
 		&				& \ in \Bc 					& {330\ }	& {396\ }	& \\
 		& \ SNR $\ge 3 $		& \ in \Sc 					& {214\ }	& {316\ }	& \\
 		&				& \ in \Hc 					& {189\ }	& {213\ }	& 							\\ \hline
 		& 				&$<$400$''$ for ACIS-I Obs.,			& 		& 	& 								\\
\up{level 3}	& \up{Off-axis limit}		& CCD 7 for ACIS-S	Obs.			& \up{276\ }	& \up{242\ }	& \up{518\ \ }	&& \up{16\x2}	& \up{2} & \up{12}	\\ \hhline{~----}
 		&				& \ in \Bc 					& {214\ }	& {210\ } 	& \\
 		& \ SNR $\ge 3 $		& \ in \Sc 					& {131\ }	& {169\ } 	& \\
 		&				& \ in \Hc 					& {123\ } 	& {110\ } 	& \\
\hline
\end{tabular}
\end{center}
Notes.~--- See Table~\ref{t:flag} for flag definitions.\\
\sS{a}Special attention required: sources found in multiple
observations of the same field (flag=157), potential pile-ups (flag=147), and source
region overlap (flag=141). No source with flag=142 (severe source
region overlap) is found in these fields\\ 
\sS{b}Removes extended sources and other apparently false sources. For
the case of sources near a chip boundary (flag=146), they may be valid
sources, but we do not include them for level $\ge$ 1 analysis because
their position and source properties are subject to large uncertainties. \\
\end{table*}

In the following, we go over the basic X-ray properties of the above
sources. The complete source catalog 
as well as access to the X-ray and optical images and optical 
photometry data for probable IDs  (see Grindlay et al 2005 for examples) is available 
on-line.\footnote{http://hea-www.harvard.edu/ChaMPlane/.}

\subsection{Source distribution}

We derive the \lnls distribution to review the analysis procedure and 
for an initial analysis of 
the nature of the source population.  In general, making an
estimate of the unabsorbed flux for sources in the Galactic Plane fields
is more difficult than for high-latitude sources because of the diverse
and unknown spectral types, the higher column density and its strong
dependence on the usually unknown distances to the Galactic sources.
Therefore, the derivation and interpretation of the \lnls distributions
for \champlane sources are not trivial.  For the initial analysis,
we follow a practice similar to the one done for \champ (K04), which
allows us to verify the consistency of the analysis and to compare the
results directly with the high-latitude \champ results.  We generate
the sky coverage and the \lnls distribution in the \Sc and \Hc band.
To simplify the analysis, we use only the level 3 data set.  We also
exclude the original target (and point sources in the target region if the
intended target contains diffuse emission or a cluster of X-ray sources;
flag=154) of each observation to maintain the serendipitous nature of
the survey. For example, in the NGC 1569 field (a galaxy, Obs.~ID 782),
we consider the point sources within the 3.6$'$ diameter\footnote{
Optical major diameter ($D_{25}$). See \citet{Humphrey03}.}
circle from the galaxy center as a part of
target,  and in the 3C 123 field (a quasar, Obs.~ID
829), two sources nearby 3C 123 are also considered so.

\subsubsection{Sky Coverage}

To calculate the amount of sky covered, we adopt a relatively simple
but accurate procedure \citep{Cappelluti05}.  First, we generate a
background-only image from event files for a given band.  Using the
valid source list (level 1), we remove the photons in the source
regions and fill the regions with counts consistent
with the neighboring background using {\it
dmfilth}.\footnote{http://cxc.harvard.edu/ciao/ahelp/dmfilth.html}  Note
we keep the photons in the source region of the target (and sources in
the target region if the target is 
extended) 
in the background-only image
because we exclude these sources in counting the number of sources for
\lnls.  Second, for a given
position in a CCD, using the exposure map and the background image, we
estimate the count rate limit ($r_l$) for detection at a given
signal-to-noise ratio (SNR$=\sigma$) by
\begin{eqnarray}
	r_l = \sigma \frac{1 +(0.75 + N_B)^{1/2}}{ T}
		\frac{\mbox{max}(e_p)}{\overline{e_p}|_{p \in A_B}},
	\label{e:rl}
\end{eqnarray}
where $N_B$ is the background count
in the source region, $T$ the exposure time, $\sigma$ = 3 and
we use Gehrels' approximation for the errors \citep{Gehrels86}.\footnote{
The Gehrels' approximation in Eq.~(\ref{e:rl}) is optimal for $\sigma =1$, and
for $\sigma > 1$, \citet{Gehrels86} recommands a more sophisticated
formula, but we take the above simpler approach to be consistent with
the definition of SNR in Eq.~(\ref{e:snr}).}

We need to convert $r_l$ to the equivalent flux in order to calculate the
sky coverage as a function of flux.  Note that $r_l$ is scaled by the
exposure map in the same way as $r\Ss{net}$ in Eq.~(\ref{e:rate}).  Therefore,
one can use the same rate-to-flux conversion factor in the previous
section (\S\ref{s:flux}) to convert $r_l$ to the equivalent flux.  However,
there is no unique conversion from the count rate to un-absorbed
flux because sources in the Galaxy can be at any distance from the
Earth and the column density changes with the distance.  As for direct
comparison with the \champ results, we again assume the full column
density along the line of the sight provided by the extinction models.
Since \citet{Schlegel98} may overestimate and \citet{Drimmel03} may
underestimate the total extinction along the line of the sight, we use
both models (insets in Fig.~\ref{f:lnls}).

As for spectral models, we employ a power-law emission model with
$\PLI=1.7$ for the \Sc band and $\PLI=1.4$ for the
\Hc band  sources as in \champ (K04). These assumptions for spectral
and column density models are more relevant for background AGN sources
than for Galactic sources.  

Using the rate-to-flux conversion factor and Eq.~(\ref{e:rl}), one can
assign the flux limit for $\sigma=3$ at any position in the detector. For
a given flux, the sky coverage is derived by summing up the detector
area where the calculated flux limit is less than the given flux. In
summation, we only include the level 3 region where exposure map is
greater than 10\% of the maximum of the CCD to meet the source selection
criterion (flag $\ne$ 146).

\subsubsection{\lnls}

To be consistent with the sky coverage calculation ($\sigma\ge3$), for a
given band, we count the number of sources with SNR$ \ge 3$ for the \lnls
distribution except for the target and sources in the target region, where 
\begin{eqnarray}
 	\mbox{SNR} = \frac{N\Ss{net}}{1 +(0.75 + N\Ss{src}-N\Ss{net})^{1/2}}.
	\label{e:snr}
\end{eqnarray}

\begin{figure*} \begin{center} 
\epsscale{0.57}
\plotone{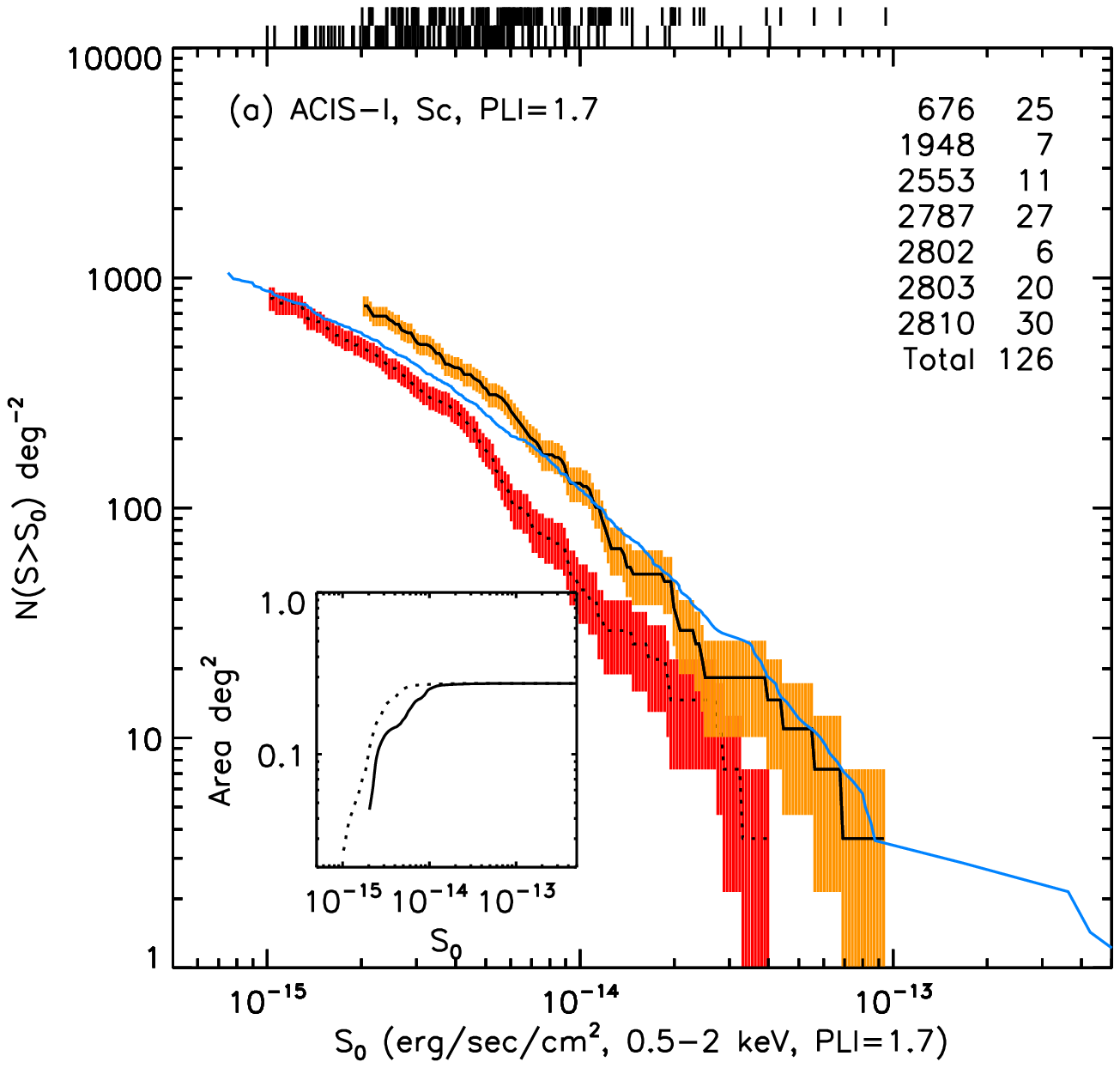} 
\plotone{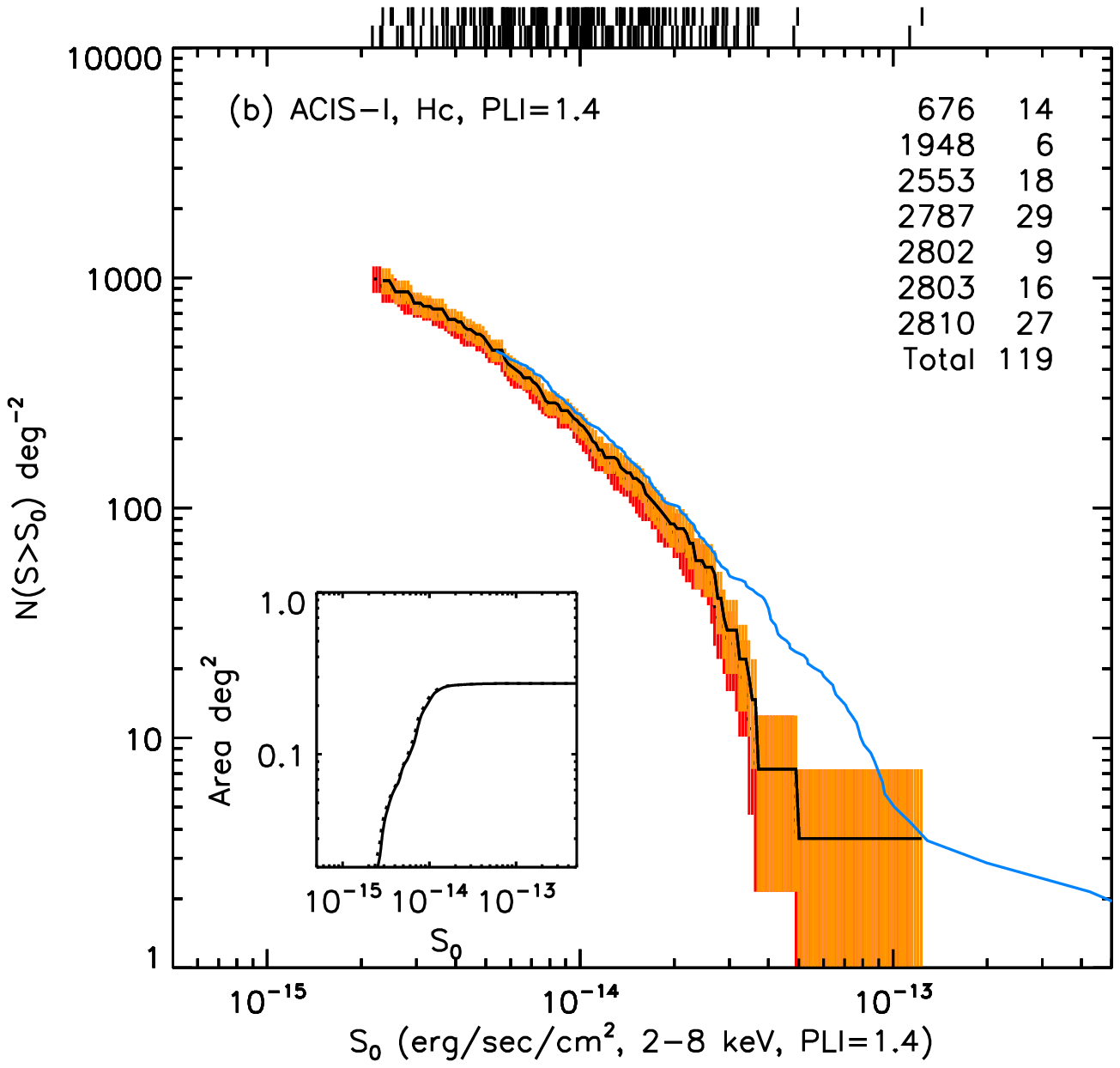} 
\plotone{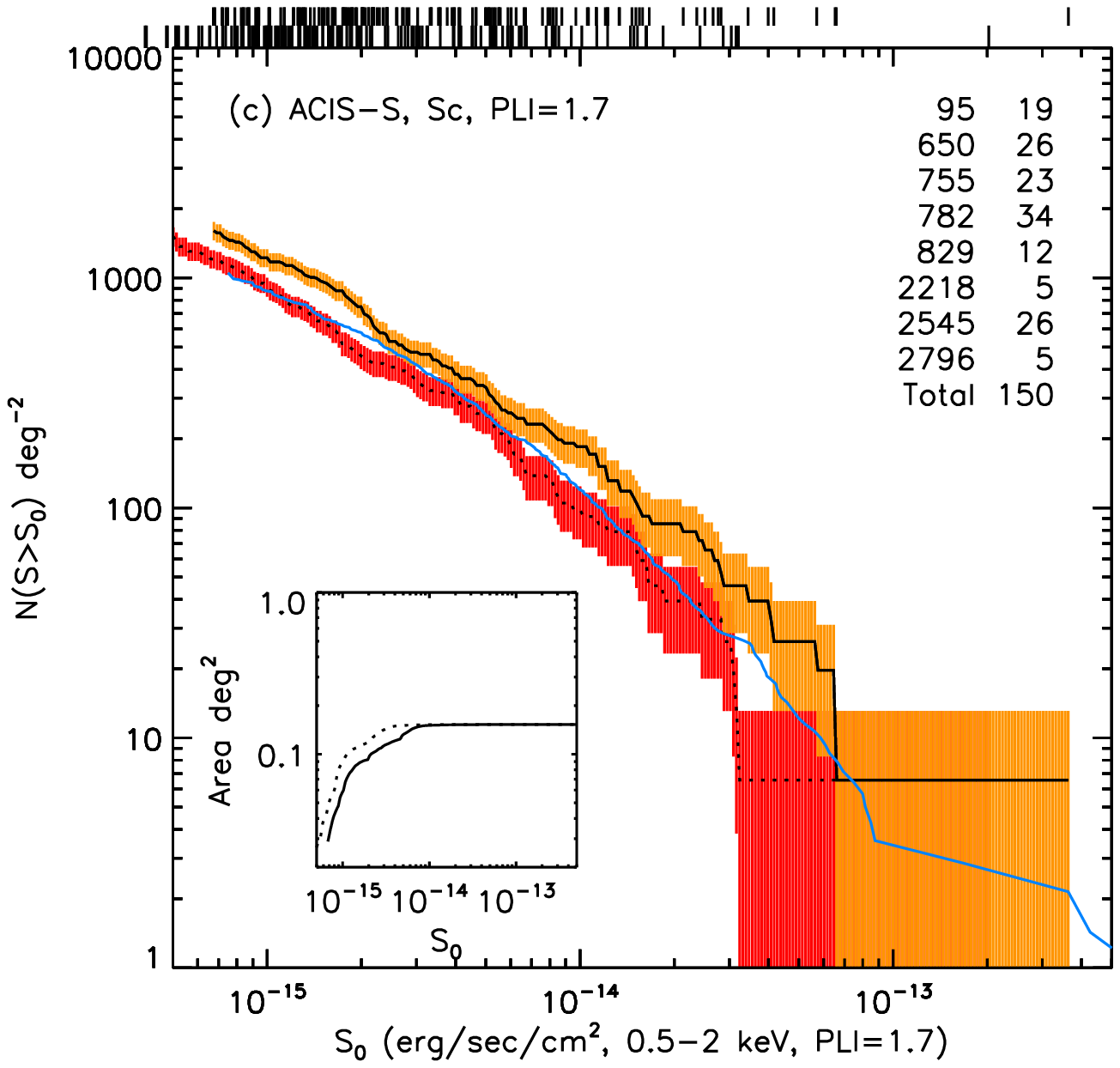} 
\plotone{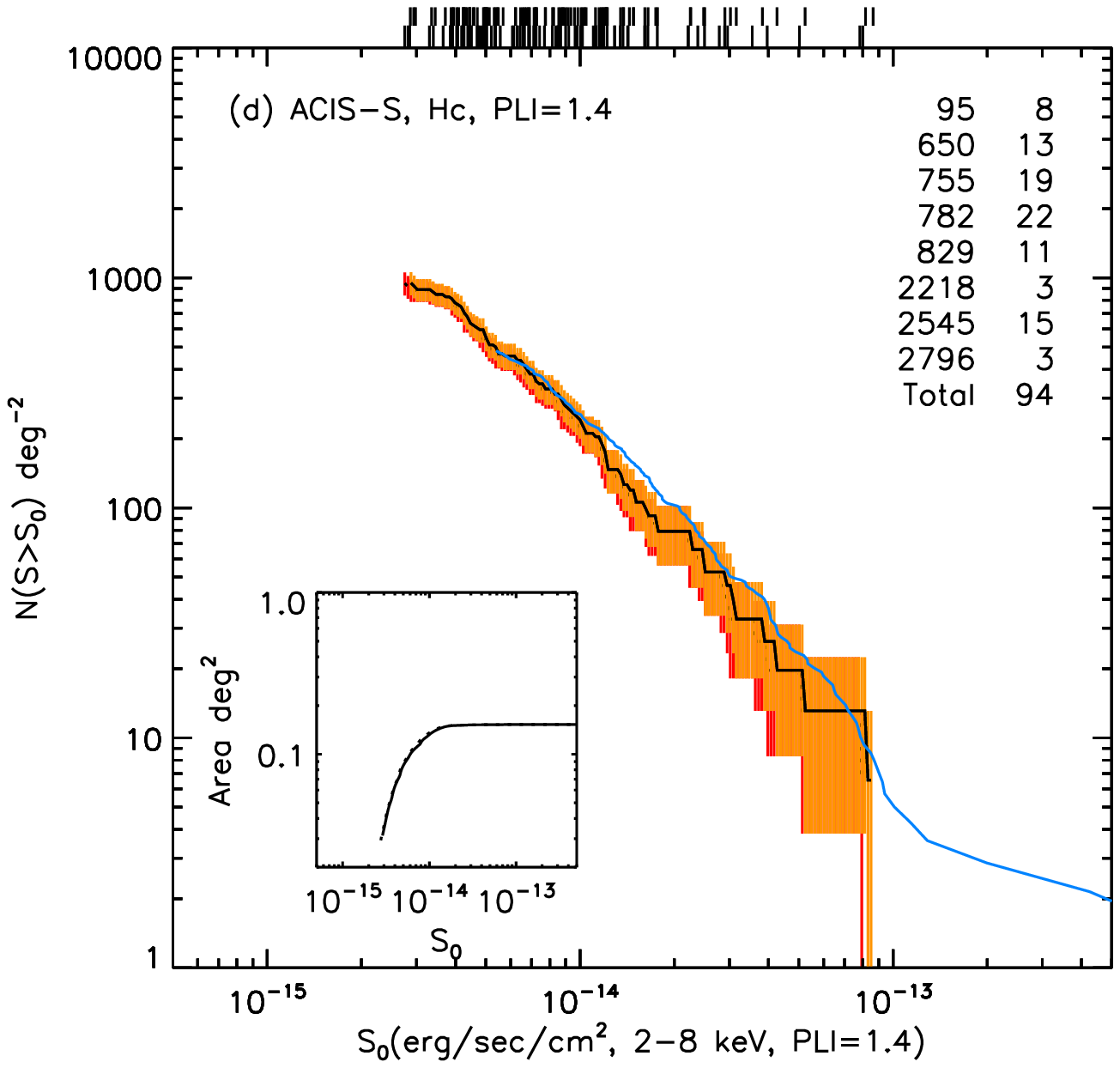} 
\end{center} 
\caption{The \lnls distributions of the X-ray sources in the \noO
anti-GC observations ($90\Deg < l < 270\Deg$): 
(a) ACIS-I observations in the \Sc band, (b) ACIS-I in \Hc, 
(c) ACIS-S in \Sc,
and (d) ACIS-S in \Hc.
The solid line with the (yellow) shade is for the extinction model by
\citet{Schlegel98}, and the dotted line with the (red) shade is for the model by
\citet{Drimmel03}. The shading represents the
statistical uncertainties ($1 \sigma$) of the distribution. The thin
(blue) solid line is the extra-Galactic result
from \champ (K04). The inset shows the sky coverage: the solid line
from the model by \citet{Schlegel98}, and the dotted line from \citet{Drimmel03}.
The top right list in each plot indicates the Obs.~IDs and the number of sources
with SNR $\ge 3$. The targets of each observation are excluded. 
The vertical black bars above each plot show the actual source
distribution by flux assuming the model by \citet{Schlegel98} (top row) and
the flux distribution assuming the model by \citet{Drimmel03} (bottom rows).
See the
electronic ApJ for color version of the plots.} 
\label{f:lnls} 
\end{figure*}

Fig.~\ref{f:lnls} shows the \lnls distribution of the sources in the
anti-GC fields.  The solid line with the (yellow) shade is for the
extinction model by \citet{Schlegel98}, and the dotted line with the
(red) shade is for the model by \citet{Drimmel03}. The shading represents
the statistical uncertainties ($\pm1 \sigma$) of the distribution. The
\champ results are shown as a thin (blue) solid line without any shade
in the plots.  Table~\ref{t:lnls} summarizes the current
status of source classification (only $\sim$ 6 -- 17\%) by optical
imaging and spectrocopy for the sources in Fig.~\ref{f:lnls}. 

In Fig.~\ref{f:lnls}, for a given spectral and extinction model, the
results for the ACIS-I and ACIS-S
observations agree within $\le 2 \sigma$ overall, and they exhibit
better agreement ($\le 1 \sigma$) where the 
number of sources is sufficient so that the statistical fluctuations are
not too overwhelming ($S_0< 1
\times 10^{-14}$ erg/sec/cm\sS{2} in the \Sc band and $S_0< 3 \times
10^{-14}$ erg/sec/cm\sS{2} in the \Hc band). 

Because of the relatively poor statistics, it is premature to
attribute differences between the ACIS-I and ACIS-S observations to
any systematic bias in the analysis.  For example, in the case of the
\Sc band [Fig.~\ref{f:lnls} (a) \& (c)], the \lnls distributions
expectedly show a strong dependence on the extinction models, which
differ by as much as a factor of 3 (Table~\ref{t:list}).  For Galactic
populations, the extinction depends strongly on the distance to the
sources. In addition, the assumed power law emission model (\PLI=1.7)
may not properly describe the X-ray emission from these sources
(\S\ref{s:qccd}). For soft coronal sources (\S\ref{s:qccd}), the assumed
power law emission model likely underestimates the source flux and thus
their contribution to the \lnls distribution, particularly in the soft
band (shifting the \lnls distribution to the left in the plot).

The unknown spectral model and the inaccurate extinction estimate
dominate the uncertainty of the \lnls distributions in the \Sc band.
Nonetheless, we expect our anti-GC results to exhibit an excess relative
to the \champ \lnls
distribution.  We suspect the extinction model of \citet{Drimmel03}
underestimates the total Galactic \nH values for at least a subset of
the fields.

In the case of the \Hc band [Fig.~\ref{f:lnls} (b) \& (d)], the \lnls
distributions for the anti-GC fields are not sensitive to the extinction
models, and the results are consistent with the \champ \lnls distribution
within the statistical limit.  This indicates that most of the strong hard
sources are background AGNs. For example, judging from the statistical
uncertainty of the \lnls distributions, we estimate that
for $S> S_0 = 2\x 10^{-14}$ erg/sec/cm\sS{2}, only less than $\sim$ 20\% of
the sources are Galactic in Fig.~\ref{f:lnls} (b) \& (d).  

It is premature to draw any conclusion from
source classification (Table~\ref{t:lnls} or Table~\ref{t:match})
because only $\sim 10$\% of the sources in Fig.~\ref{f:lnls} (b) \& (d)
are identified as of this writing.
However, the above \lnls results are consistent with source classification
in Table~\ref{t:lnls} because, among 8 sources 
(out of 15 classified sources in the \Hc band) that have $S> 2\x
10^{-14}$ erg/sec/cm\sS{2} and SNR $\ge 3$ in the \Hc band, 7 sources
are quasars and only one is Galactic.\footnote{Note the \lnls
distributions in Fig.~\ref{f:lnls} (and Table~\ref{t:lnls}) use level
3 data and exclude the target of each observation. On the other hand,
Table~\ref{t:match} does not exclude the target of observations, but 
uses selection criteria different from those in Table~\ref{t:lnls}.
Table~\ref{t:match} contains 10 sources (out of 25 distinct sources) 
that have $S> 2\x 10^{-14}$ erg/sec/cm\sS{2} and SNR $\ge 3$ in the
\Hc band, and among them, 8 sources are quasars and the other two are Galactic.}
Note that in the above comparison, we need a limit on $S$ to alleviate
the selection effects arising from spectral difference between stars
and quasars in Table~\ref{t:lnls}.

Assuming the extinction model by \citet{Schlegel98} is correct or at
least does not underestimate the total Galactic \nH value, let us
compare the results using \citet{Schlegel98} in Fig.~\ref{f:lnls}.
The \lnls distributions in the \Sc band shows a larger excess compared
to the \champ result than the distributions in the \Hc band.  Several
reasons may explain these differences.  First, the X-ray emission of many
Galactic sources, e.g. coronally active stars -- that are detected in
\champlane but not in \champ -- is relatively stronger in the soft band.
Although only $\lesssim$ 15\% of the X-ray sources
are classified (Table~\ref{t:lnls}), the results suggest that,
as expected, the star to quasar ratio in the  \lnls distribution of
the soft band is higher than the same for the hard band.  Second, for
Galactic sources, the extinction and hence the unabsorbed flux is likely
overestimated by
the model of \citet{Schlegel98} which can systematically shift the
entire curve in Fig.~\ref{f:lnls} to the right. The effects of
extinction are relatively larger in the soft than in the hard
band.

\begin{table}
\small
\begin{center}
\caption{Classification of Sources in
Fig.~\ref{f:lnls} by Optical Imaging and Spectroscopy \label{t:lnls}} 
\begin{tabular}{rrrrrrrrrr}
\hline\hline
				& \multicolumn{4}{c}{ACIS-I}	&& \multicolumn{4}{c}{ACIS-S} \\
\hhline{~----~----}
				&\multicolumn{2}{l}{(a) \Sc} 	&\multicolumn{2}{l}{(b) \Hc}&&\multicolumn{2}{l}{(c) \Sc}&\multicolumn{2}{l}{(d) \Hc} 	\\
\hline
Total				& 126	&	& 119	&	&& 150	&	& 94	&	\\
\sS{a}Matching performed 	& 89	&	& 76	&	&& 124	&	& 79	&	\\
\sS{b}Optical Matches		& 65 	&(6)	& 38 	& (5)	&& 71 	& (3)	& 33	& (0)\\
\hline
\sS{c}Spectrum Observed		& 35	&	& 23	&	&& 41	&	& 19	&	\\
Identified as Stars		& 15	&	& 5	&	&& 24	&	& 3	&	\\
Quasars				&  8	&	& 4	&	&& 3	&	& 3	&	\\
Galaxy				&  1	&	& 0	&	&& 0	&	& 0	&	\\
Unclear				& 11 	&	& 14	&	&& 14	&	&13	&	\\
\hline
\end{tabular}
\end{center}
Notes. --- The \lnls distributions in Fig.~\ref{f:lnls} exclude targets of each observation.\\
\sS{a}As of this writing, source classification with optical spectroscopies is
performed for 4 ACIS-I Obs.~and 6 ACIS-S Obs.~among total 15 Obs. 
See also \cite{Zhao05}.\\
\sS{b}() for the number of sources with multiple matching candidates. \\
\sS{c}See \citet{Rogel05}.
\end{table}

To derive the true Galactic \lnls distribution, we need to allow for
the differing source spectra and distances (and thus \nH), and then
re-derive the flux estimate of each source separately.  In future papers,
we shall incorporate additional information (e.g.~from optical spectra
and thus reddening) for sources whenever available for more complete
\lnls distributions.

\subsection{Source Properties by QCCD} \label{s:qccd}

\begin{figure*} \begin{center} 
\epsscale{0.57}
\plotone{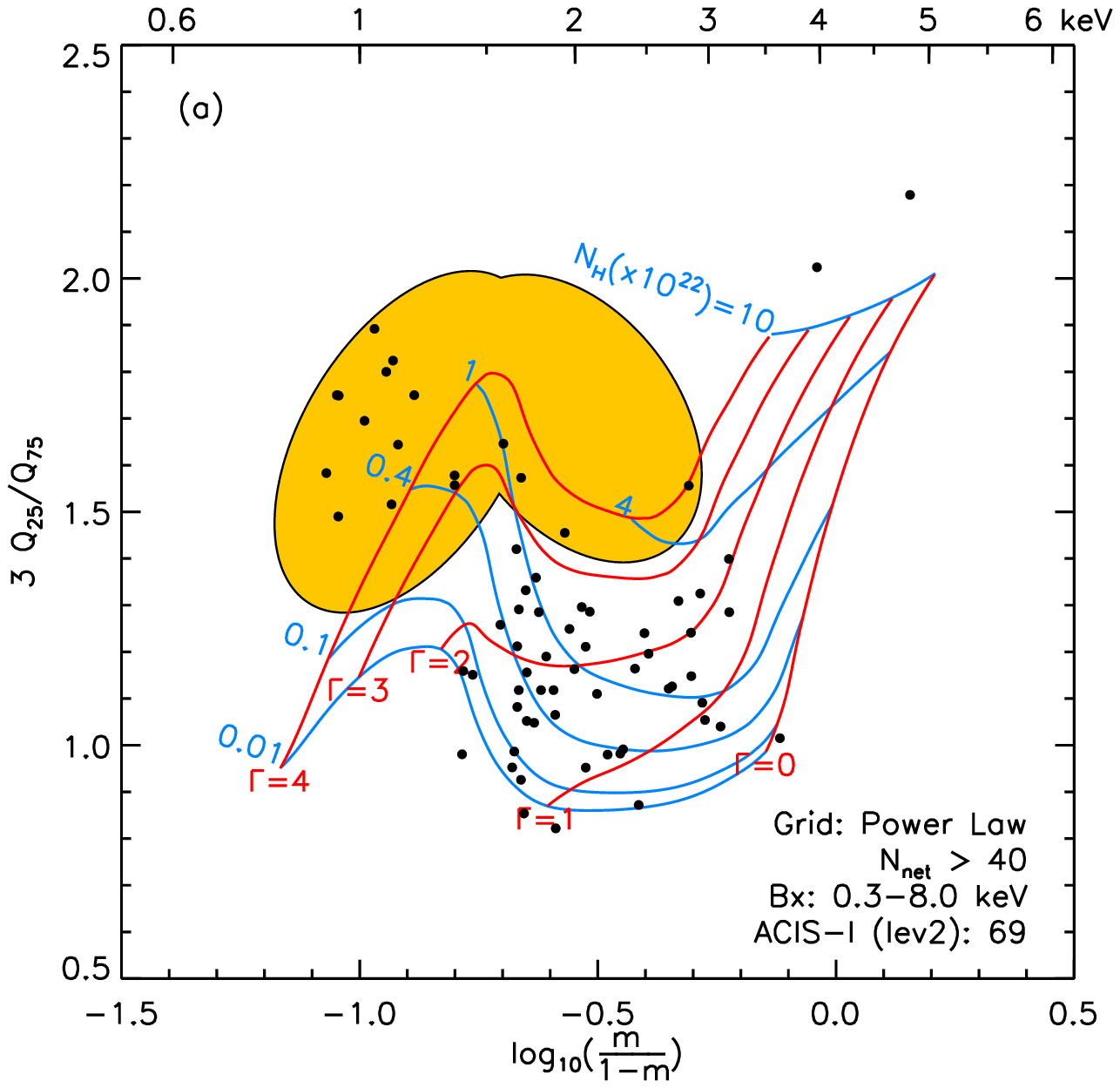}
\plotone{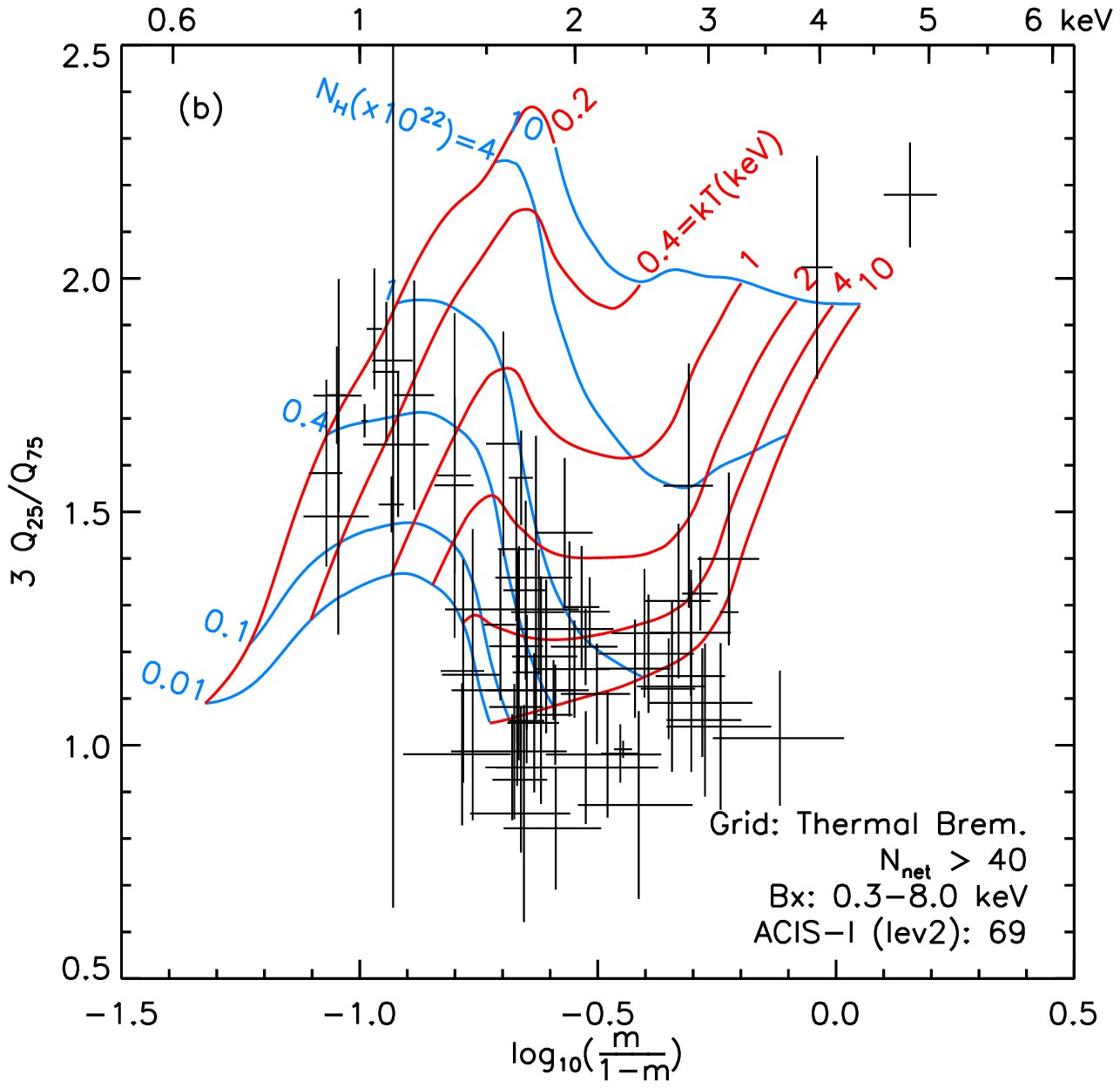}
\plotone{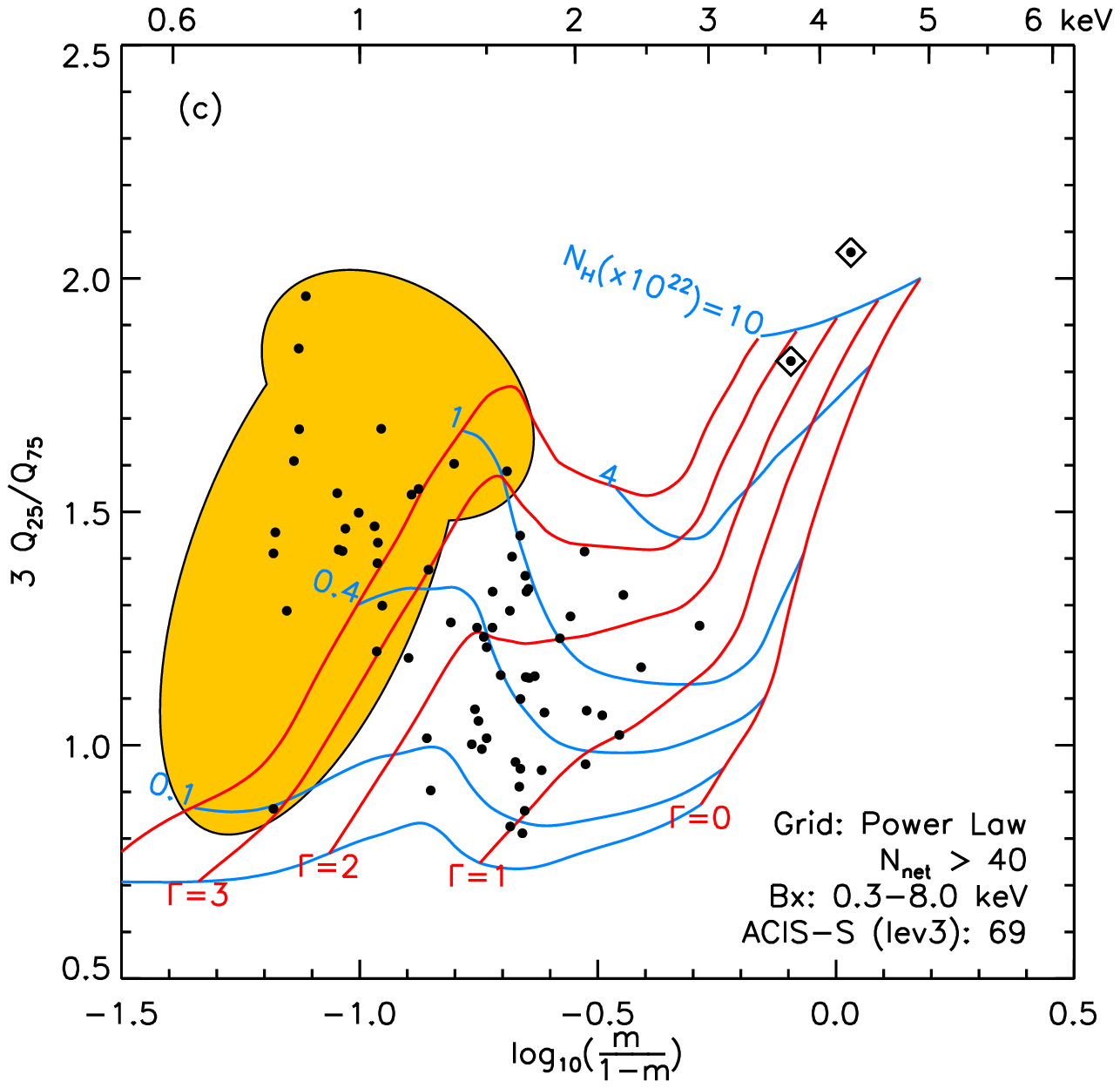} 
\plotone{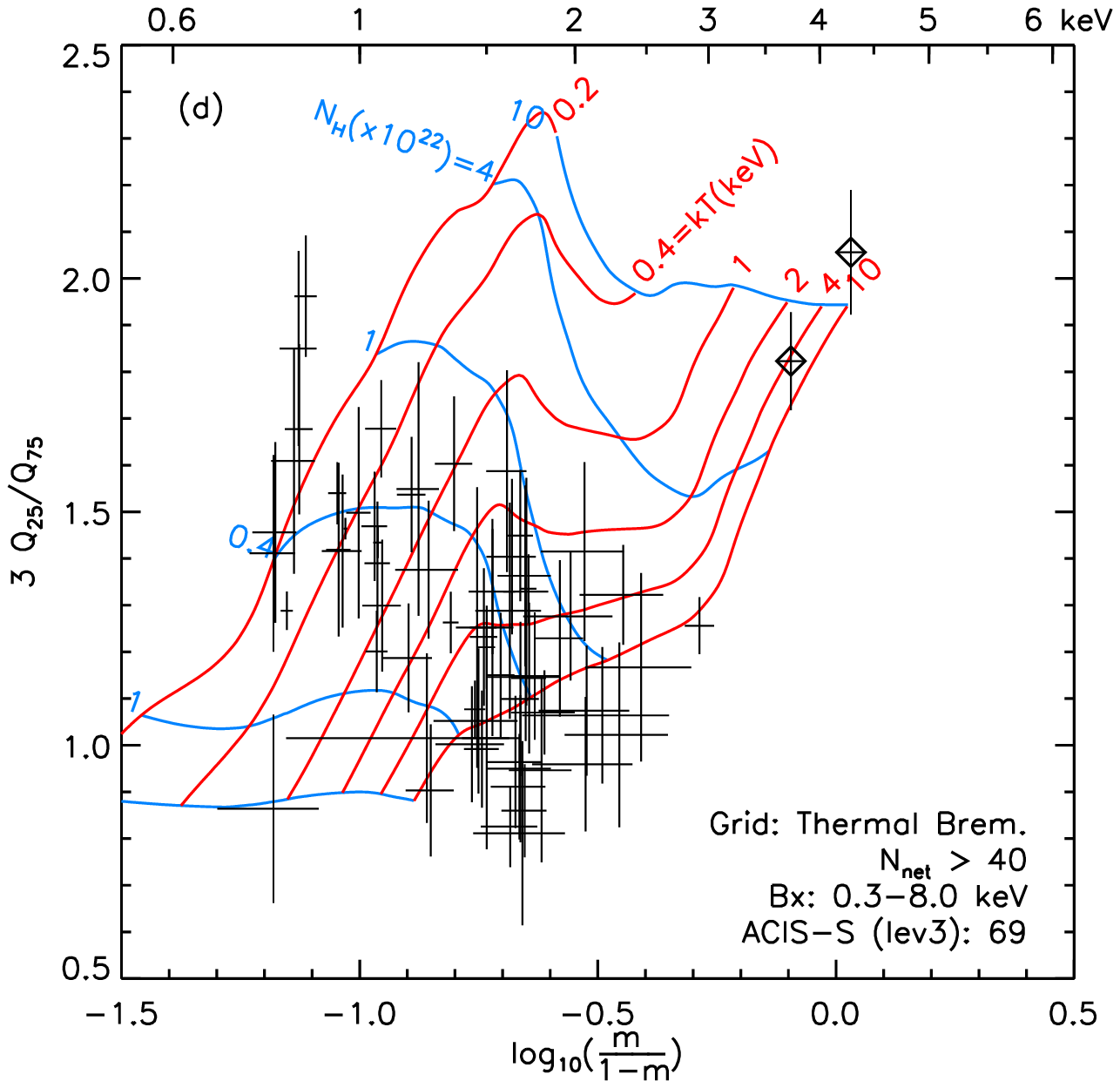} 
\end{center} 
\caption{QCCDs for the anti-GC sources ($N\Ss{net} >
40 $ in \Bx, including targets): (a) ACIS-I observations, overlayed with power law model grids,
(b) ACIS-I with error bars and thermal Brems.~model grids,
(c) ACIS-S, overlayed with power law model grids,
and (d) ACIS-S, with error bars and thermal Brems.~model grids.
The power law model grids are for $\PLI$ = 0, 1, 2, 3 \& 4 and
\nH = $10^{20}, 10^{21}, 4 \times 10^{21}, 10^{22}, 4 \times 10^{22}$
\& $10^{23}$ cm\sS{-2}, and the thermal Brems.~model grids for $kT$ =
0.2, 0.4, 1, 2, 4 \& 10
keV and the same \nH. The quantiles are calculated in the \Bx band.
The ACIS-I observations contain
sources from CCD 0, 1, 2, \& 3 (level 2, FI), and the ACIS-S from CCD 7
(level 3, BI, see Table~\ref{t:st}) except for the piled-up sources. 
In the left panels, sources in the 
shaded(yellow) regions  
are too soft to be described by power-law with physically reasonable 
\PLI and are likely dominated by stellar coronal emission sources (stars).
Diamonds in (c) and (d) indicate the two example sources in the right
panel of Fig.~\ref{f:ap};
XS00782B\_001 and XS00782B\_002.
See electronic ApJ for color version of the figure.   }
\label{f:qccd} \end{figure*}

Fig.~\ref{f:qccd} shows the QCCDs of the anti-GC field sources with $N\Ss{net}
> 40 $ in \Bx.  
The detailed description of the QCCD is found in \citet{Hong04}.
In summary, the $x$-axis shows 
the median of the photon energy distribution ($m$)
and the $y$-axis shows the ratio of
two quartiles of the photon energy distribution (3 \Qt{25}/\Qt{75}).
Note that the median $m$ is related to the
median energy ($E_{50\%}$) by
\begin{eqnarray}
	m = \frac{E_{50\%} - E\Ss{lo}}{E\Ss{up}-E\Ss{lo}}, 
	\label{e:m}
\end{eqnarray}
where $E\Ss{lo}$ = 0.3 and $E\Ss{up}$= 8.0 keV in the \Bx band \citep{Hong04}. 
The top $x$-axis of the QCCD is labeled by $E_{50\%}$ and the bottom
$x$-axis is the inverse 
hyperbolic tangent of $m$ (this allows a convenient quantity for plotting).

Fig.~\ref{f:qccd} (a) is for ACIS-I observations, overlayed with power-law
model grids, and Fig.~\ref{f:qccd} (b) shows the same data with the
error bars and thermal Bremsstrahlung model grids. Fig.~\ref{f:qccd}
(c) and (d) are for ACIS-S observations.  The figures contain
sources from 
level 2 data for ACIS-I, and level 3 data for ACIS-S observations 
(see Table~\ref{t:st}) 
except for two piled-up sources
(PSR J0538+2817 and GK Persei).  Note the difference in the grid
patterns between the FI and BI CCDs, which is mainly due to relatively
high detection efficiency of BI CCDs at low energies.  The temporal
degradation of the low energy efficiency also makes the grid patterns
change (shrink). The grid patterns in Fig.~\ref{f:qccd} are averaged
over the selected observations.

In Fig.~\ref{f:qccd} (a) and (c), one can roughly identify sources in
a few distinct groups.  In particular, the sources 
within the shaded (yellow) region 
appear to be too soft to be described by power-law models ($\PLI
\gtrsim 3 - 4$). The rest are relatively well described by
power law models with conventional values for $\PLI$ ($\sim$ 1 -- 3)
but reveal a wide range of extinctions.  In Fig.~\ref{f:qccd} (b) and (d),
one can see the very soft sources are better described by thermal models
($kT \lesssim 1 $ keV), which indicates many of these sources are most
likely to be stellar coronal emission sources (stars).  The hard sources
that are better described by the power law models are likely accreting
X-ray sources and are predominantly background AGN but can also include
accretion-powered compact binaries (e.g.~cataclysmic variables; see
Grindlay et al 2005).

\begin{figure*} 
\epsscale{0.57}
\plotone{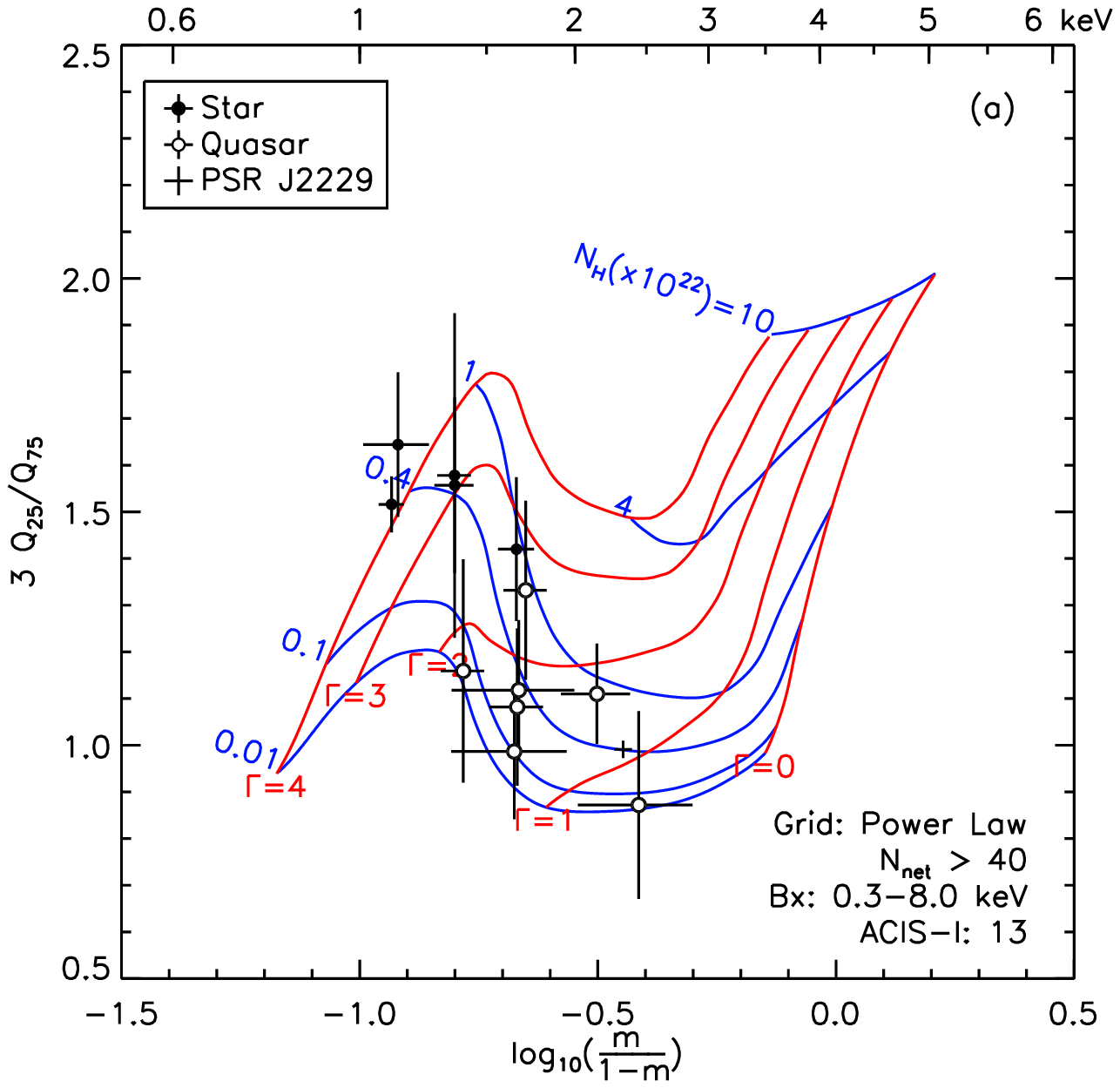} 
\plotone{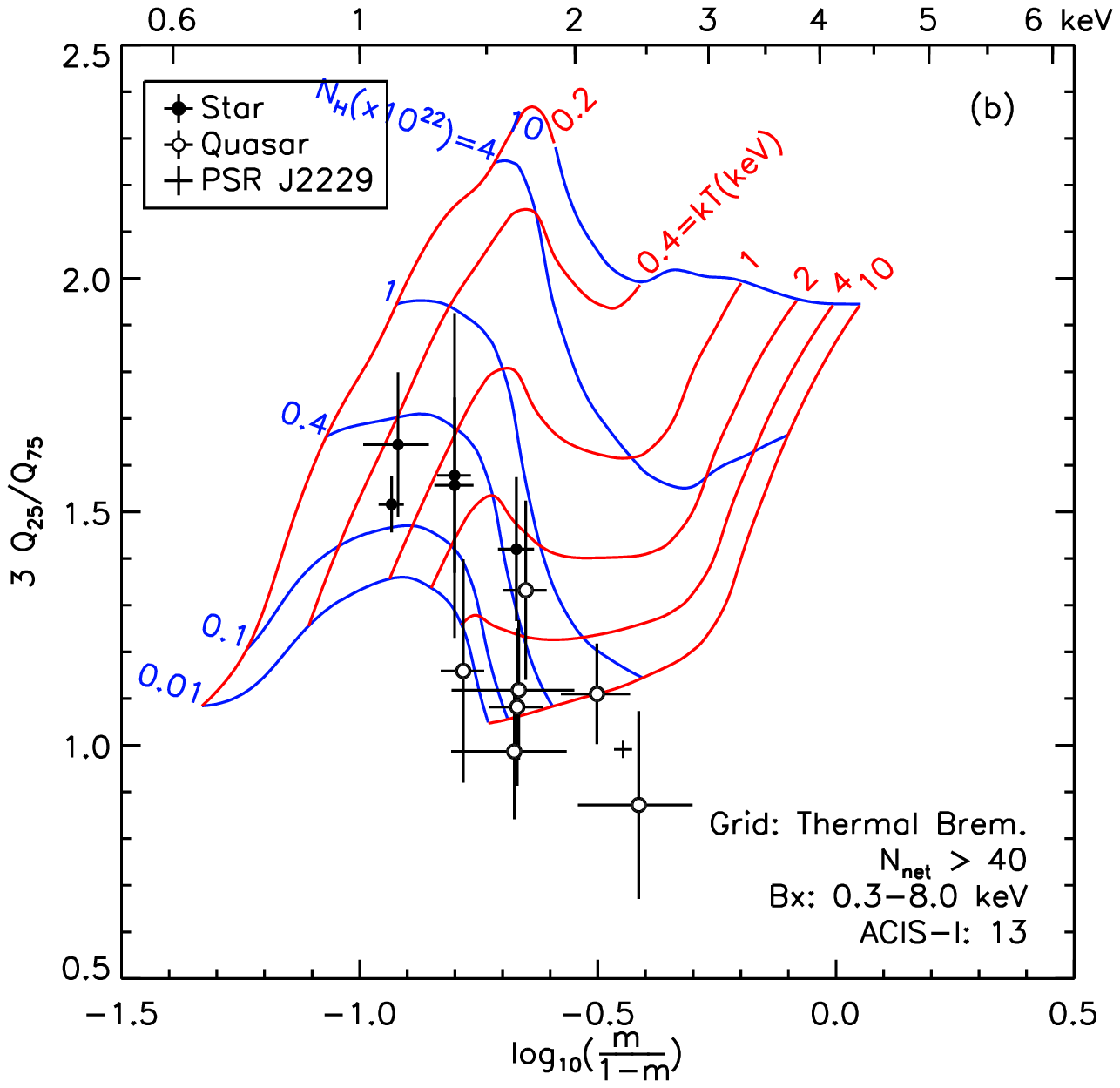} 
\plotone{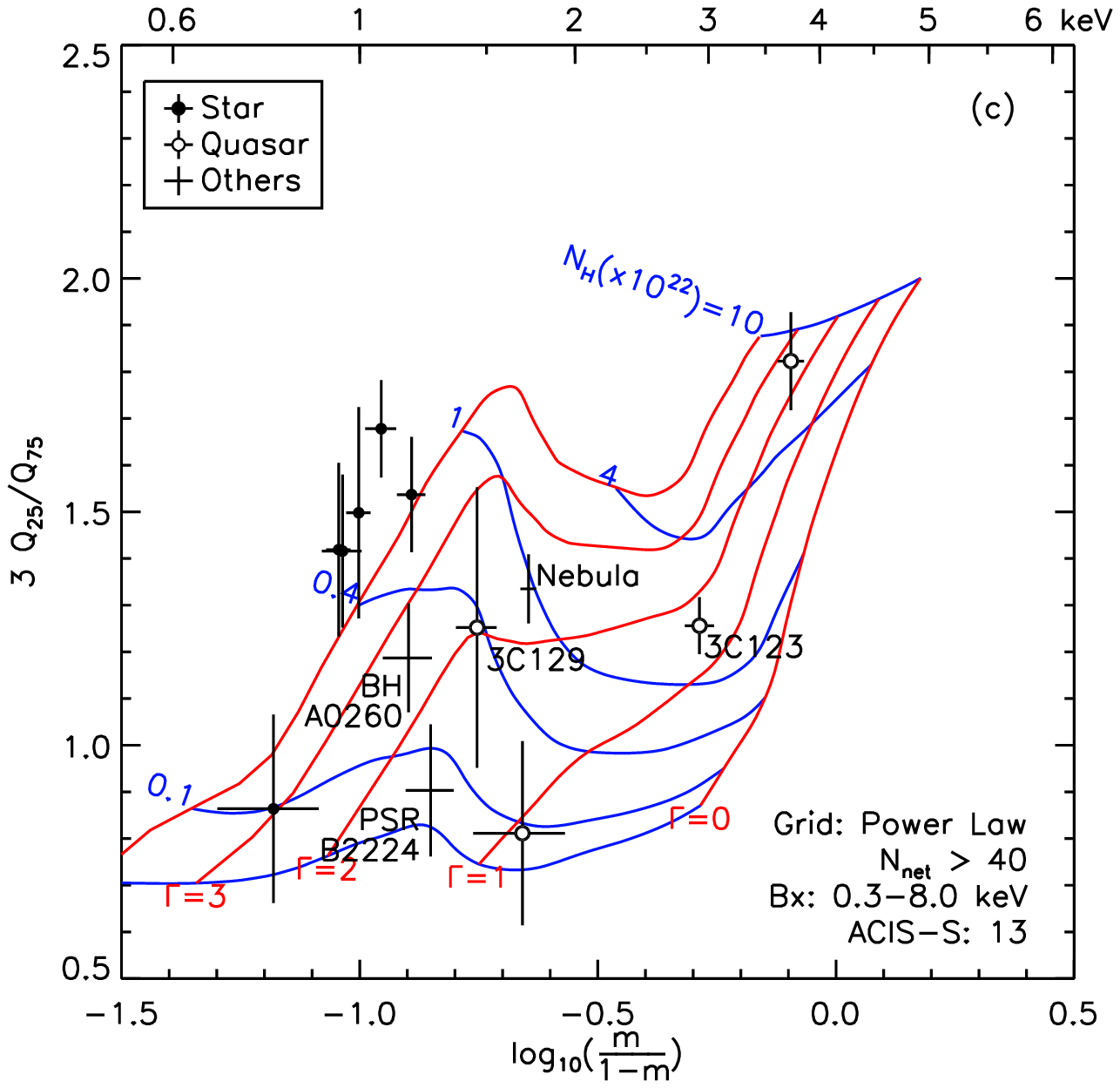} 
\plotone{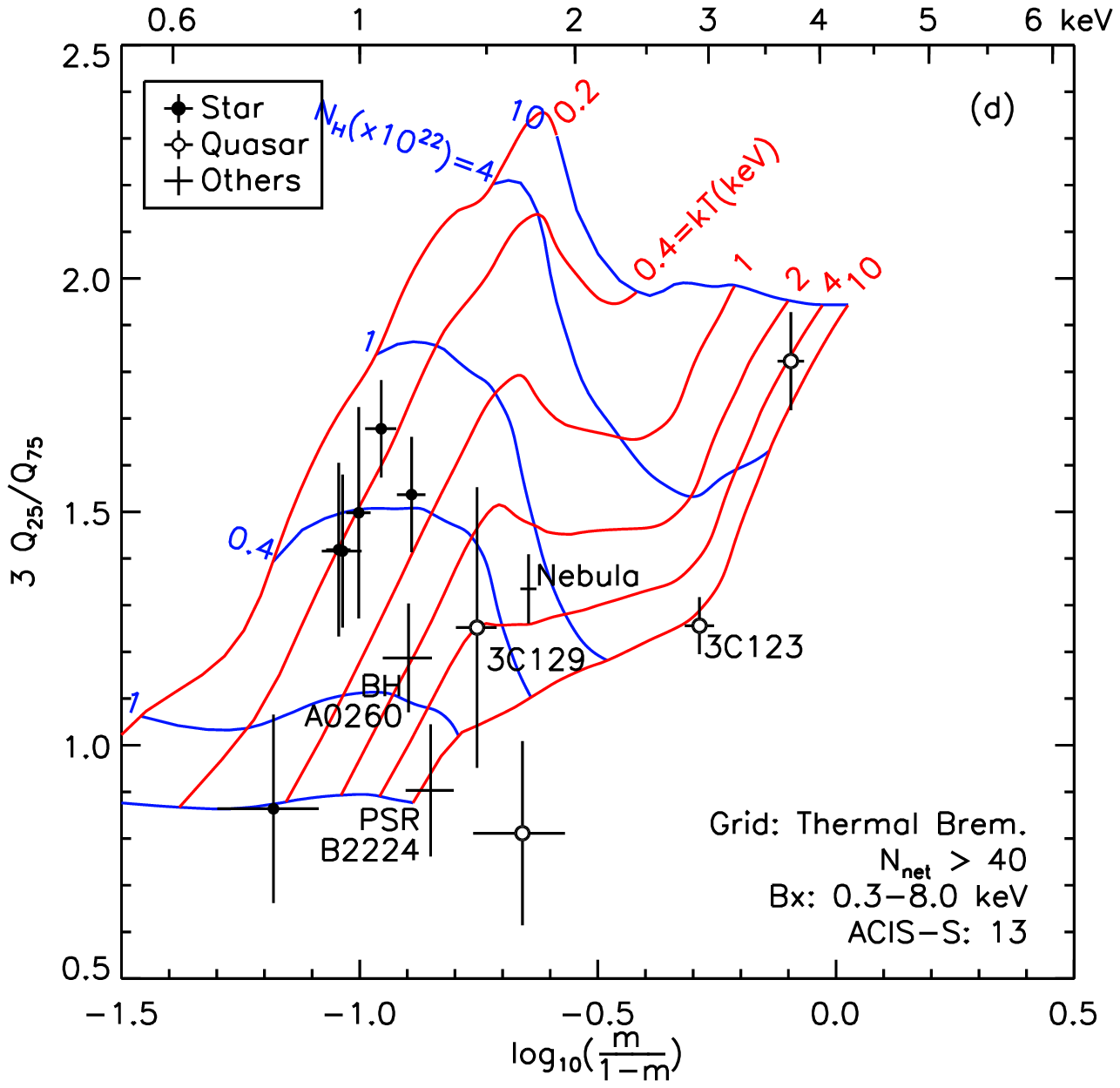} 
\caption{The same QCCDs as in Fig.~\ref{f:qccd} but only with the optically
identified sources \citep{Rogel05} or known sources ($N\Ss{net} > 40$ in \Bx). 
The solid circles are stars and the open
circles are background quasars and the rest are black hole binaries, pulsars, etc. See
Table \ref{t:match} for the complete source list of the figure. 
}
\label{f:match} 
\end{figure*}

\begin{table*}
\small
\begin{center}
\caption{List of optically identified or known sources with $N\Ss{net} > 40$ in \Bx
\label{t:match}}
\begin{tabular}{r@{}c@{\hspace{3mm}}c@{\hspace{3mm}}c@{\hspace{3mm}}cr@{\hspace{3mm}}r@{}l@{\hspace{3mm}}lc}
\hline\hline
&Source ID		&Source Name  	&$E_{50\%}$\sS{a}& 				& $N\Ss{net}\ $ & \multicolumn{2}{@{}c}{$r\Ss{net}$ in \Bx}		& 			& 	\\
&XS			&CXOPS	 	&(keV)		 &\up{$3\Qt{25}/\Qt{75}$}	& in \Bx 	& \multicolumn{2}{@{}c}{(cts/ksec)}		& \up{Source Type}	& \up{Known Name\sS{b}}	\\
\hline
\multicolumn{3}{l}{\it CCD 0,1,2,3 (FI) for ACIS-I observations} \\
\hspace{10mm} 
	&02810B2\_018 	&J235813.4$+$622447  		& 2.14(23)  	&1.11(11)   	&  97.7 	& 2.4&  (3)& Quasar z=1.29	\\ 
	&00676B3\_011	&J042155.1$+$324724  		& 1.71(12)  	&1.33(19)   	&  64.2 	& 3.8&  (5)& Quasar z=1.845\\
	&00676B0\_001 	&J042201.6$+$325728  		& 1.39(10) 	&1.16(24)   	&  63.5 	& 3.7&  (5)& Quasar z=2.203\\
	&00676B3\_018 	&J042123.7$+$324836  		& 1.66(14)  	&1.08(17)   	&  57.0 	& 3.6&  (5)& Quasar 	\\ 
	&00676B2\_001	&J042133.8$+$325556  		& 1.64(30) 	&0.99(14)   	&  48.5 	& 2.7&  (5)& Quasar z=2.055\\
  \sS{c}&02810B0\_021	&J235813.2$+$623343  		& 2.44(42)  	&0.87(20)   	&  42.4 	& 1.0&  (2)& Quasar z=2.04	\\	
	&00676B1\_006 	&J042211.8$+$325604  		& 1.67(33)  	&1.12(15)   	&  41.8 	& 2.7&  (5)& Quasar z=0.65	\\ \hhline{~---------}
	&02787B3\_002	&J222905.2$+$611409  		& 2.33(06)  	&0.99(02) 	&1972.6		&21.9&  (5)&  \\
  \sS{d}&01948B3\_001 	&J222905.2$+$611409  		& 2.31(13)  	&0.98(06) 	& 327.4		&25.2&  (1.5)& \up{Pulsar}		&\up{PSR J2229+6114}	\\ \hhline{~---------}
	&02787B1\_005 	&J222833.4$+$611105  		& 1.10(04)  	&1.52(06)  	& 237.9 	& 2.9&  (2)& \\ 
  	&01948B1\_002 	&J222833.3$+$611105  		& 1.13(12)  	&1.64(16)   	&  46.8 	& 3.5&  (6)& \up{YSO K star}		\\ \hhline{~---------} 
	&02787B0\_003	&J223001.4$+$611059  		& 1.35(09)  	&1.56(19)   	&  87.6 	& 1.1&  (1)& \\ 
  \sS{e}&01948B0\_001	&J223001.3$+$611100  		& 1.62(34)  	&1.41(76)   	&  16.8 	& 1.3&  (4)& \up{early K star}	\\ \hhline{~---------}
	&02787B3\_007  	&J222847.3$+$611214   		& 1.65(10)      &1.42(15)    	&  79.1  	& 1.6&  (2)&      \\
  \sS{e}&01948B1\_001	&J222847.3$+$611214   		& 1.76(21)   	&1.45(26)    	&  25.4   	& 1.8&  (4)& \up{G star}      \\ \hhline{~---------}
	&02787B3\_005 	&J222853.1$+$611351  		& 1.35(07)  	&1.58(35)   	&  48.0 	& 0.5&  (1)& \\ 
  \sS{e}&01948B3\_004	&J222853.0$+$611351  		& 1.40(39)  	&1.38(37)   	&   8.4 	& 0.6&  (3)& \up{G/K? star}	\\ \hline
\multicolumn{4}{l}{\it CCD 7 (BI) for ACIS-S observations} \\
	&00829B7\_002	&J043704.3$+$294013  		&2.92(12)  	& 1.26(06) 	&  380.7 	&14.1&  (8)&Quasar z=0.218  & 3C 123\\
	&00782B7\_002	&J043125.0$+$645154  		&3.73(12)  	& 1.82(10) 	&  157.8 	& 2.7&  (2)&Quasar z=0.279\\ 
	&00829B7\_003	&J043654.8$+$294018  		&1.69(25)  	& 0.81(20) 	&   45.1 	& 1.7&  (3)&Quasar 	\\
	&02218B7\_004	&J044909.0$+$450039  		&1.45(10)  	& 1.25(30) 	&   47.2 	& 2.3&  (4)&Quasar	  & 3C 129 \\ \hhline{~---------}
	&00829B7\_022  	&J043718.4$+$294546		&1.07(05)   	& 1.68(10)   	&  136.2  	& 7.4&  (7)&K star \\  
	&00782B7\_042	&J043003.0$+$645143  		&1.18(05)  	& 1.54(12) 	&  133.6 	& 1.7&  (2)&YSO M star 	\\ 
	&00650B7\_015	&J033108.2$+$435750  		&0.94(03)  	& 1.42(19) 	&  111.3 	& 2.0&  (2)&dMe star 	\\
	&00755B7\_005	&J222545.8$+$653833  		&0.95(06)  	& 1.42(16) 	&   75.4 	& 1.7&  (2)&dMe star 	\\
	&00782B7\_010	&J043058.3$+$644852  		&1.00(04)  	& 1.50(23) 	&   63.7 	& 0.8&  (1)&dMe star 	\\
	&00782B7\_018	&J043039.0$+$645013  		&0.78(11)  	& 0.86(20) 	&   40.2 	& 0.5&  (1)&dMe star 	\\ \hhline{~---------}
	&00782B7\_012	&J043057.4$+$645048  		&1.72(04)  	& 1.34(07) 	&  257.8 	& 3.1&  (2)&Nebula  \\ 
	&00095B7\_004	&J062244.5$-$002044  		&1.17(09)  	& 1.19(12) 	&   99.7 	& 3.1&  (4)&BH XB 	  & A0620-00 \\
	&00755B7\_004	&J222552.5$+$653535  		&1.25(10)  	& 0.90(14) 	&   83.7 	& 1.8&  (2)&Pulsar 	  & B2224+65\\
\hline
\end{tabular}\\
\end{center}
\sS{a} See Eq.~(\ref{e:m}).  The top $x$-axis of the QCCDs is labeled by $E_{50\%}$. \\
\sS{b} Ones with the known name is the target of the observation. \\
\sS{c} Three candidates are found for the counter part.  The nearest candidate is
identified as a quasar and the other two are unknown type.  \\
\sS{d} This source is not plotted in Fig.~\ref{f:match} for clarity. \\
\sS{e} Note $N\Ss{net} < 40$ for these sources (not shown in Fig.~\ref{f:match}), 
	but they are listed for comparison 
	with the same sources found in Obs.~ID 2787 with  $N\Ss{net} > 40$. \\
\end{table*}

Fig.~\ref{f:match} shows the same QCCD plots but only with the known
sources or optically identified ones \citep{Rogel05}. Table \ref{t:match}
contains the complete list of the sources in Fig.~\ref{f:match}. The solid
circles indicate stars and the open circles indicate background AGNs.
The rest are pulsars, black hole (BH) X-ray binary (XB), etc.  The QCCD
separates relatively soft stars and hard quasars even with a small
number of net counts ($>$ 40 counts in \Bx). Five sources in Obs.~ID
2787 are also found in Obs.~ID 1948. Note that their quantile values
(and net count rates) from two observations are consistent, even though
three sources in Obs.~ID 1948 have only 16.8, 25.4 and 8.4 net counts
respectively (Table~\ref{t:match}).

\section{Summary}

We describe the X-ray analysis procedure 
for X-ray processing of ChaMPlane survey archival data 
using custom developed analysis tools, which can be useful for other 
Chandra analysis projects as well.  
The initial
X-ray results from the analysis of the \noO selected anti-GC observations
reveal a few distinct classes of sources.  The detailed X-ray
analysis of source distributions and source classifications  
for all Anticenter ChaMPlane fields, followed by Galactic Bulge and GC
region fields, will be presented in subsequent papers.

\section{Acknowledgements}
We thank Terrance Gaetz for performing the SAOSAC 
and MARX simulations. We also thank Dong-woo Kim, Minsun
Kim and Eunhyeuk Kim for the useful discussion and suggestions, and
we appreciate help from the \champ team for \champlane data
processing/analysis.
This work is supported in part by NASA/Chandra grants AR1-2001X, AR2-3002A, 
AR3-4002A, AR4-5003A and NSF grant AST-0098683.


\begin{thebibliography}{}

\bibitem[Baganoff et al.(2003)]{Baganoff03}  
	Baganoff, F.K.~et al., 2003,
	ApJ, 591, 891.

\bibitem[Bohlin et al.(1978)]{Bohlin78}  
 	Bohlin, R.C., Savage, B.D. \& Drake, J.F.,  1978,
	ApJ, 224, 132.
 
\bibitem[Cappelluti et al.(2005)]{Cappelluti05}
   	Cappelluti, N. et al., 2005,
	ApJ, 430, 39.


\bibitem[Drimmel et al.(2003)]{Drimmel03}  
	Drimmel, R., Cabrera-Lavers, A., \&
	Lopez-Corredoira, M., 2003, 
	ApJ, 409, 205.

\bibitem[Freeman et al.(2002)]{Freeman02}  
	Freeman, P.E.~et al., 2002,
	ApJS, 138, 185.

\bibitem[Gehrels et al.(1986)]{Gehrels86}  
	Gehrels, N.~et al., 1986,
	ApJ, 303, 336.

\bibitem[Grindlay et al.(2003)]{Grindlay03}  
	Grindlay, J.E.~et al., 2003,
	AN, 324, 57.

\bibitem[Grindlay et al.(2005)]{Grindlay05}  
	Grindlay, J.E.~et al., 2005,
	submitted to ApJ.

\bibitem[Humphrey et al.(2003)]{Humphrey03}
	Humphrey, P.J.~et al., 2003,
	MNRAS, 344, 134.

\bibitem[Hong et al.(2004)]{Hong04}  
 	Hong, J, Schlegel, E.M.~\& Grindlay, J.E., 2004,
	ApJ, 614, 508.

\bibitem[Kim et al.(2004a)]{Kim04a}  
	Kim, D.-W.~et al., 2004a,
	ApJS, 150, 19 (K04).

\bibitem[(2004b)]{Kim04b}  
	Kim, D.-W.~et al., 2004b,
	ApJ, 600, 59 (K04).

\bibitem[Kim et al.(2005)]{Kim05}  
	Kim, E., 2005,
	in preparation, private communication.

\bibitem[Laycock et al.(2005)]{Laycock05}  
	Laycock, S.G.T.~et al., 2005,
	in preparation.

\bibitem[Mallat,~S.(1998)]{Mallat98}
	Mallat, S.~1998, 
	A Wavelet Tour of Signal Processing (London: Academic Press).

\bibitem[Predehl \& Schmitt(1995)]{Predehl95}
 	Predehl, P. \& Schmitt, J. H. M. M.,
	1995, A\&A, 293, 889.

\bibitem[Rogel et al.(2005)]{Rogel05}  
	Rogel, A.~et al., 2005,
	submitted to ApJ. 

\bibitem[Schlegel et al.(1998)]{Schlegel98}  
	Schlegel, D., Finkbeiner, D., \& Davis, M.,
  	ApJ, 1998, 500, 525.

\bibitem[Zhao et al.(2003)]{Zhao03}  
	Zhao, P.~et al., 2003,
	AN, 324, 176.

\bibitem[Zhao et al.(2005)]{Zhao05}  
	Zhao, P.~et al., 2005,
	submitted to ApJ.
	
\end{thebibliography}
\end{document}